\documentclass[twocolumn,preprintnumbers,amsmath,amssymb]{revtex4}

\usepackage{graphicx}
\usepackage{dcolumn}
\usepackage{bm}
\usepackage{rotating}

\usepackage{epstopdf}
\usepackage{bbm}
\usepackage{verbatim}
\usepackage{txfonts}
\usepackage[cspex,bbgreekl]{mathbbol}

\usepackage{color}
\definecolor{lightblue}{rgb}{0.2,0.2,0.7}
\definecolor{darkblue}{rgb}{0,0.25,0.5}
\definecolor{redbrown}{rgb}{0.875,0.25,0.125}
\definecolor{darkgreen}{rgb}{0,0.5,0}

\newcommand{\bra}[1]{\ensuremath{\langle #1 \vert}}
\newcommand{\ket}[1]{\ensuremath{\vert #1  \rangle}}

\renewcommand{\b}[1]{\ensuremath{\mathbf{#1}}}
\renewcommand{\H}{\ensuremath{\text{H}}}
\renewcommand{\l}{\ensuremath{\lambda}}

\newcommand{\lr}{\ensuremath{\text{lr}}}
\newcommand{\sr}{\ensuremath{\text{sr}}}

\newcommand{\HF}{\ensuremath{\text{HF}}}
\newcommand{\IP}{\ensuremath{\text{IP}}}
\newcommand{\CBS}{\ensuremath{\text{CBS}}}

\newcommand{\RSH}{\ensuremath{\text{RSH}}}

\newcommand{\T}{\ensuremath{\text{T}}}
\DeclareMathOperator{\erf}{erf}

\begin{document}

\title{Range-separated density-functional theory with random phase approximation:\\ detailed formalism and illustrative applications}

\author{Julien Toulouse$^{1}$}\email{julien.toulouse@upmc.fr}
\author{Wuming Zhu$^{1}$}\email{wuming@lct.jussieu.fr}
\author{J\'anos G. \'Angy\'an$^{2}$}\email{janos.angyan@crm2.uhp-nancy.fr}
\author{Andreas Savin$^{1}$}\email{savin@lct.jussieu.fr}
\affiliation{
$^1$ Laboratoire de Chimie Th\'eorique, UPMC Univ Paris 06 and CNRS, 75005 Paris, France\\
$^2$ CRM2, Institut Jean Barriol, Nancy University and CNRS, 54506 Vandoeuvre-l\`{e}s-Nancy, France
}


\date{\today}
\begin{abstract}
Using Green-function many-body theory, we present the details of a formally exact adiabatic-connection fluctuation-dissipation density-functional theory based on range separation, which was sketched in Toulouse, Gerber, Jansen, Savin and \'Angy\'an, Phys. Rev. Lett. {\bf 102}, 096404 (2009). Range-separated density-functional theory approaches combining short-range density functional approximations with long-range random phase approximations (RPA) are then obtained as well-identified approximations on the long-range Green-function self-energy. Range-separated RPA-type schemes with or without long-range Hartree-Fock exchange response kernel are assessed on rare-gas and alkaline-earth dimers, and compared to range-separated second-order perturbation theory and range-separated coupled-cluster theory.
\end{abstract}

\maketitle

\section{Introduction}

Range-separated density-functional theory has emerged as a powerful approach for improving the accuracy of standard Kohn-Sham (KS) density-functional theory~\cite{HohKoh-PR-64,KohSha-PR-65} applied with usual local or semi-local density-functional approximations, in particular for electronic systems with strong (static) or weak (van der Waals) correlation effects. Based on a separation of the electron-electron interaction into long-range and short-range components, it permits a rigorous combination of a long-range explicit many-body approximation with a short-range density-functional approximation (see, e.g., Ref.~\onlinecite{TouColSav-PRA-04} and references therein). Several many-body approximations have been considered for the long-range part: configuration interaction~\cite{LeiStoWerSav-CPL-97,PolSavLeiSto-JCP-02}, multi-configuration self-consistent-field theory~\cite{PedJen-JJJ-XX,FroTouJen-JCP-07,FroReaWahWahJen-JCP-09}, second-order perturbation theory~\cite{AngGerSavTou-PRA-05,GerAng-CPL-05b,GerAng-JCP-07,GolLeiManMitWerSto-PCCP-08,JanScu-PCCP-09}, coupled-cluster theory~\cite{GolWerSto-PCCP-05,GolWerStoLeiGorSav-CP-06,GolStoThiSch-PRA-07,GolWerSto-CP-08,GolErnMoeSto-JCP-09}, multi-reference second-order perturbation theory~\cite{FroCimJen-PRA-10}, and several variants of the random phase approximation (RPA)~\cite{TouGerJanSavAng-PRL-09,JanHenScu-JCP-09,JanHenScu-JCP-09b,JanScu-JCP-09,PaiJanHenScuGruKre-JCP-10}.

In the context of the recent revived interest in RPA-type approaches to the electron correlation problem in atomic, molecular and solid-state systems~\cite{YanPerKur-PRB-00,Fur-PRB-01,AryMiyTer-PRL-02,MiyAryKotSchUsuTer-PRB-02,FucGon-PRB-02,NiqGon-PRB-04,FucNiqGonBur-JCP-05,FurVoo-JCP-05,DahLeeBar-PRA-06,MarGarRub-PRL-06,JiaEng-JCP-07,HarKre-PRB-08,Fur-JCP-08,ScuHenSor-JCP-08,RenRinSch-PRB-09,LuLiRocGal-PRL-09,HarKre-PRL-09,NguGir-PRB-09,NguGal-JCP-10,GruMarHarSchKre-JCP-09,HelBar-JCP-10,HarSchKre-PRB-10,Ism-PRB-10,RuzPerCso-JCTC-10}, several range-separated approaches using long-range RPA-type approximations have indeed been proposed and show promising results, in particular for describing weak intermolecular interactions. Toulouse {\it et al.}~\cite{TouGerJanSavAng-PRL-09} have presented a range-separated RPA-type theory including the long-range Hartree-Fock exchange response kernel. Janesko {\it et al.}~\cite{JanHenScu-JCP-09,JanHenScu-JCP-09b,JanScu-JCP-09} have proposed a simpler range-separated RPA scheme with no exchange kernel and in which the RPA correlation energy has been rescaled by an empirical coefficient. Paier {\it et al.}~\cite{PaiJanHenScuGruKre-JCP-10} have added the so-called second-order screened exchange to the latter scheme, which appears to correct the self-interaction error. In all these cases, range separation tends to improve the corresponding full-range RPA-type approach, avoiding the inaccurate description and slow basis-set convergence of short-range correlations in RPA.

In Ref.~\onlinecite{TouGerJanSavAng-PRL-09}, only the main lines of range-separated density-functional theory with long-range RPA were presented. In this work, we give now all the missing details of the theory. Using Green-function many-body theory, we construct a formally exact adiabatic-connection fluctuation-dissipation density-functional theory based on range separation, without the need of maintaining the one-particle density constant. Range-separated RPA-type schemes are then obtained as well-identified approximations on the long-range Green-function self-energy. The range-separated RPA-type methods with or without long-range Hartree-Fock exchange response kernel are assessed on rare-gas and alkaline-earth dimers, and compared to range-separated second-order perturbation theory and range-separated coupled-cluster theory. The most tedious details of the theory are given in the appendices.

\section{Theory}

\subsection{Range-separated density-functional theory}

In range-separated density-functional theory (see, e.g., Ref.~\onlinecite{TouColSav-PRA-04}), the exact ground-state energy of an $N$-electron system is expressed as the following minimization over multideterminant wave functions $\Psi$
\begin{equation}
E  = \min_{\Psi} \left\{ \bra{\Psi} \hat{T} + \hat{V}_{ne} + \hat{W}_{ee}^{\lr} \ket{\Psi} + E_{\H xc}^{\sr}[n_{\Psi}]\right\},
\label{EminPsi}
\end{equation}
where $\hat{T}$ is the kinetic energy operator, $\hat{V}_{ne}$ is the nuclei-electron interaction operator, $\hat{W}_{ee}^{\lr} \!=\! (1/2) \iint d\b{r}_1 d\b{r}_2 w_{ee}^{\lr}(r_{12}) \hat{n}_2(\b{r}_1,\b{r}_2)$ is a long-range (lr) electron-electron interaction written with $w_{ee}^{\lr}(r)\! =\!\erf(\mu r)/r$ and the pair-density operator $\hat{n}_2(\b{r}_1,\b{r}_2)$, and $E_{\H xc}^{\sr}[n]$ is the corresponding $\mu$-dependent short-range (sr) Hartree-exchange-correlation (Hxc) density functional that Eq.~(\ref{EminPsi}) defines. The parameter $\mu$ in the error function controls the range of the separation. The minimizing wave function, denoted by $\Psi^\lr$, yields the exact density. Several approximations~\cite{TouSavFla-IJQC-04,TouColSav-PRA-04,TouColSav-JCP-05,GolWerSto-PCCP-05,PazMorGorBac-PRB-06,FroTouJen-JCP-07,GolErnMoeSto-JCP-09} have been proposed for the short-range exchange-correlation (xc) functional $E_{xc}^{\sr}[n]$, and an approximate scheme must be used for the long-range wave function part of the calculation.

In a first step, the minimization in Eq.~(\ref{EminPsi}) is restricted to single-determinant wave functions $\Phi$, leading to the range-separated hybrid (RSH) approximation~\cite{AngGerSavTou-PRA-05} 
\begin{equation}
E_{\RSH}  = \min_{\Phi} \left\{ \bra{\Phi} \hat{T} + \hat{V}_{ne} + \hat{W}_{ee}^{\lr} \ket{\Phi} + E_{\H xc}^{\sr}[n_{\Phi}]\right\},
\label{ERSHminPhi}
\end{equation}
which does not include long-range correlation. The minimizing determinant $\Phi_0$ is given by the self-consistent Euler-Lagrange equation
\begin{equation}
\hat{H}_0 \ket{\Phi_0} = {\cal E}_{0} \ket{\Phi_0},
\label{H0Phi0}
\end{equation}
where ${\cal E}_{0}$ is the Lagrange multiplier for the normalization constraint and $\hat{H}_0$ is the RSH reference Hamiltonian
\begin{equation}
\hat{H}_0 = \hat{T} + \hat{V}_{ne} + \hat{V}_{\H x,\HF}^{\lr}[\Phi_0] +  \hat{V}_{\H xc}^{\sr}[n_{\Phi_0}],
\label{H0}
\end{equation}
which includes the Hartree-Fock (HF)-type long-range Hartree-exchange (Hx) potential $\hat{V}_{\H x,\HF}^{\lr}[\Phi_0]$ and the short-range local Hxc potential $\hat{V}_{\H xc}^{\sr}[n] \!=\! \int d\b{r} v_{\H xc}^{\sr}[n](\b{r}) \hat{n}(\b{r})$ written with $v_{\H xc}^{\sr}[n](\b{r}) = \delta E_{\H xc}^{\sr}[n]/\delta n(\b{r})$ and the density operator $\hat{n}(\b{r})$. As usual, $\hat{V}_{\H x,\HF}^{\lr}$ is the sum of a local Hartree part $\hat{V}^{\lr}_{\H}=\int d\b{r}_1 v_\H^{\lr}(\b{r}_1) \hat{n}(\b{r}_1)$ with $v_\H^{\lr}(\b{r}_1)=\int d\b{r}_2 w_{ee}^{\lr}(r_{12}) \bra{\Phi_0} \hat{n}(\b{r}_2) \ket{\Phi_0}$, and a non-local exchange part $\hat{V}^{\lr}_{x,\HF}=\iint d\b{x}_1 d\b{x}_2 v_x^{\lr}(\b{x}_1,\b{x}_2) \hat{n}_1(\b{x}_2,\b{x}_1)$ written with $v_x^{\lr}(\b{x}_1,\b{x}_2)=- w_{ee}^{\lr}(r_{12}) \bra{\Phi_0} \hat{n}_1(\b{x}_1,\b{x}_2) \ket{\Phi_0}$ and the one-particle density-matrix operator $\hat{n}_1(\b{x}_1,\b{x}_2)$ expressed with space-spin coordinates $\b{x}_1=(\b{r}_1,s_1)$ and $\b{x}_2=(\b{r}_2,s_2)$.

The RSH scheme does not yield the exact energy and density, even with the exact short-range functional $E_{\H xc}^{\sr}[n]$. Nevertheless, the RSH approximation can be used as a reference to express the exact energy as
\begin{eqnarray}
E = E_{\RSH} + E_c^{\lr},
\label{ERSH+Eclr}
\end{eqnarray}
defining the long-range correlation energy $E_c^{\lr}$, for which we will now give an adiabatic connection formula. We introduce the following energy expression with a formal coupling constant $\l$
\begin{eqnarray}
E_{\l}  = \min_{\Psi} \Bigl\{ \bra{\Psi} \hat{T} + \hat{V}_{ne} + \hat{V}_{\H x,\HF}^{\lr}[\Phi_0] + \l \hat{W}^{\lr}  \ket{\Psi}
\nonumber\\
+ E_{\H xc}^{\sr}[n_{\Psi}] \Bigl\},
\label{ElminPsi}
\end{eqnarray}
where the minimization is done over multideterminant wave functions $\Psi$, $\hat{W}^{\lr}$ is the long-range M{\o}ller-Plesset-type fluctuation perturbation operator
\begin{eqnarray}
\hat{W}^{\lr} = \hat{W}_{ee}^{\lr} - \hat{V}^{\lr}_{\H x,\HF}[\Phi_0],
\label{Wlr}
\end{eqnarray}
and $E_{\H xc}^{\sr}$ is the previously-defined $\l$-independent short-range Hxc functional. The minimizing wave function, denoted by $\Psi^{\lr}_{\l}$, is given by the self-consistent Euler-Lagrange equation
\begin{eqnarray}
\hat{H}^\lr_\l \ket{\Psi^\lr_\l} =  {\cal E}^\lr_\l \ket{\Psi^\lr_\l},
\label{HlrlPsilrl}
\end{eqnarray}
where ${\cal E}^\lr_\l$ is the Lagrange multiplier for the normalization constraint and $\hat{H}^\lr_\l$ is the long-range interacting effective Hamiltonian along the adiabatic connection
\begin{eqnarray}
\hat{H}^\lr_\l &=& \hat{T} + \hat{V}_{ne} + \hat{V}_{\H x,\HF}^{\lr}[\Phi_0] + \hat{V}_{\H x c}^{\sr}[n_{\Psi^\lr_\l}] + \l \hat{W}^{\lr},
\nonumber\\
&=& \hat{H}_0 + \l \hat{W}^{\lr} + \left( \hat{V}_{\H x c}^{\sr}[n_{\Psi^\lr_\l}] - \hat{V}_{\H x c}^{\sr}[n_{\Phi_0}]\right).
\label{Hlrl}
\end{eqnarray}
For $\l=1$, Eq.~(\ref{ElminPsi}) reduces to Eq.~(\ref{EminPsi}), and so the physical energy $E=E_{\l=1}$ and density are recovered. For $\l=0$, the minimizing wave function is the RSH determinant $\Psi^{\lr}_{\l=0} = \Phi_0$ and the Hamiltonian of Eq.~(\ref{Hlrl}) reduces to the RSH reference Hamiltonian, $\hat{H}^\lr_{\l=0}=\hat{H}_0$. Note that, because the density at $\l=0$ is not exact, the density necessarily varies along this adiabatic connection.
Taking the derivative of $E_{\l}$ with respect to $\l$, noting that $E_{\l}$ is stationary with respect to $\Psi^{\lr}_{\l}$, and reintegrating between $\l=0$ and $\l=1$ gives
\begin{eqnarray}
E = E_{\l=0} + \int_{0}^{1} d\l \, \, \bra{\Psi^{\lr}_\l} \hat{W}^{\lr}  \ket{\Psi^{\lr}_\l},
\label{}
\end{eqnarray}
with $E_{\l=0}\! =\! \bra{\Phi_0} \hat{T} \!+\! \hat{V}_{ne} \!+\!  \hat{V}_{\H x,\HF}^{\lr}[\Phi_0] \ket{\Phi_0}\! +\! E_{\H xc}^{\sr}[n_{\Phi_0}]\! =\! E_{\RSH}\! -\!\bra{\Phi_0} \hat{W}^{\lr} \ket{\Phi_0}$. Thus, the long-range correlation energy is
\begin{eqnarray}
E_c^{\lr} = \int_{0}^{1} d\l \left[ \bra{\Psi^{\lr}_\l} \hat{W}^{\lr}  \ket{\Psi^{\lr}_\l} - \bra{\Phi_0} \hat{W}^{\lr}  \ket{\Phi_0} \right],
\label{}
\end{eqnarray}
or, equivalently,
\begin{eqnarray}
E_c^{\lr} = \frac{1}{2} \int_{0}^{1} d\l \int d\b{x}_1 d\b{x}_2 d\b{x}_1' d\b{x}_2' w^{\lr}(\b{x}_1,\b{x}_2;\b{x}_1',\b{x}_2') 
\nonumber\\
\times P_{c,\l}^{\lr}(\b{x}_1,\b{x}_2;\b{x}_1',\b{x}_2'),
\label{Eclr}
\end{eqnarray}
where $w^{\lr}(\b{x}_1,\b{x}_2;\b{x}_1',\b{x}_2')=w_{ee}^{\lr}(r_{12}) \delta(\b{x}_1-\b{x}_1')\delta(\b{x}_2-\b{x}_2')-1/(N-1)\left[ v_\H^{\lr}(\b{r}_1) \delta(\b{x}_1-\b{x}_1') + v_x^{\lr}(\b{x}_1,\b{x}_1') \right] \delta(\b{x}_2-\b{x}_2')$ is the potential corresponding to the perturbation operator $\hat{W}^{\lr}$ and $P_{c,\l}^{\lr}(\b{x}_1,\b{x}_2;\b{x}_1',\b{x}_2')$ is the correlation part of the two-particle density matrix along the adiabatic connection.

\subsection{Long-range many-body perturbation theory}

We now derive a formally exact many-body perturbation theory to calculate the long-range correlation two-particle density matrix $P_{c,\l}^{\lr}$. Details are given in Appendix~\ref{app:manybody}.

The one-particle Green function $G_{\l}^{\lr}(1,2)$ along the adiabatic connection of Eq.~(\ref{Hlrl}) in terms of space-spin-time coordinates $1=(\b{x}_1,t_1)$ and $2=(\b{x}_2,t_2)$ satisfies the following Dyson equation
\begin{eqnarray}
\left(G_{\l}^{\lr}\right)^{-1}(1,2) = G_{0}^{-1}(1,2) - \Sigma_{\l}^\lr(1,2) - \Delta \Sigma_{\l}^\sr(1,2),
\label{Glrlm1}
\end{eqnarray}
where $G_{0}(1,2)$ is the reference Green function corresponding to the RSH Hamiltonian $\Hat{H}_0$, $\Sigma_{\l}^\lr(1,2)$ is the self-energy corresponding to the long-range perturbation operator $\l \hat{W}^{\lr}$ and $\Delta \Sigma_{\l}^\sr(1,2)$ is the self-energy correction associated with the short-range potential variation term $\hat{V}_{\H x c}^{\sr}[n_{\Psi^\lr_\l}] - \hat{V}_{\H x c}^{\sr}[n_{\Phi_0}]$ due to the variation of the density~\cite{TouZhuAngSav-JJJ-XX-note}. The long-range self-energy corresponding to the perturbation operator $\l (\hat{W}_{ee}^{\lr} - \hat{V}^{\lr}_{\H x,\HF}[\Phi_0] )$ is decomposed into Hartree, exchange and correlation contributions as
\begin{eqnarray}
\Sigma_{\l}^\lr(1,2) &=& \Sigma_{\H x c,\l}^\lr[G^{\lr}_\l](1,2) - \Sigma_{\H x,\l}^\lr[G_0](1,2)
\nonumber\\
&=&\l \left\{\Sigma_{\H x}^\lr[G^{\lr}_\l](1,2) - \Sigma_{\H x}^\lr[G_0](1,2) \right\}
\nonumber\\
&& + \Sigma_{c,\l}^\lr[G^\lr_\l](1,2),
\label{}
\end{eqnarray}
where $\Sigma_{\H x}^{\lr}[G](1,2)$ is the sum of a long-range Hartree self-energy
\begin{eqnarray}
\Sigma_{\H}^{\lr}[G] (1,2) &=&  -i \int d3 \, d4 \, w_{ee}^{\lr}(1,3) \delta(1,2) \delta(3,4) G(4,3^+)
\nonumber\\
&=& -i \delta(1,2) \int d3 \, w_{ee}^{\lr}(1,3) G(3,3^+)
\nonumber\\
&=& \delta(1,2) \int d\b{r}_3 \, w_{ee}^{\lr}(r_{13}) n(\b{r}_3)
\nonumber\\
&=& \delta(1,2) v_\H^\lr[n](\b{r}_{1}),
\label{SigmaH}
\end{eqnarray}
with the instantaneous electron-electron interaction $w_{ee}^{\lr}(1,3)=\delta(t_1-t_3) w_{ee}^{\lr}(r_{13})$ and the density extracted from the Green function $n(\b{r}_3) = -i \sum_{s_3} G(3,3^+)$ (where $3^+$ stands for $\b{x}_3 t_3^+$ with $t_3^+=t_3+\eta$ and $\eta$ is an infinitesimal positive shift), and a long-range exchange self-energy
\begin{eqnarray}
\Sigma_{x}^{\lr}[G] (1,2) &=&  i \int d3 \, d4 \, w_{ee}^{\lr}(1,3) \delta(1,4) \delta(2,3) G(4,3^+)
\nonumber\\
&=& i w_{ee}^{\lr}(1,2) G(1,2^+)
\nonumber\\
&=& - \delta(t_1-t_2) w_{ee}^{\lr}(r_{12}) n_1(\b{x}_1,\b{x}_2)
\nonumber\\
&=& \delta(t_1-t_2) v_x^\lr[n_1](\b{x}_1,\b{x}_2),
\label{Sigmax}
\end{eqnarray}
with the one-particle density matrix extracted from the Green function $n_1(\b{x}_1,\b{x}_2) = -i G(\b{x}_1 t_1,\b{x}_2 t_1^+)$. The short-range self-energy correction corresponding to the operator $\hat{V}_{\H x c}^{\sr}[n_{\Psi^\lr_\l}] - \hat{V}_{\H x c}^{\sr}[n_{\Phi_0}]$ is written as
\begin{eqnarray}
\Delta \Sigma_{\l}^\sr(1,2) &=& \Sigma_{\H x c}^\sr[G^{\lr}_\l](1,2) - \Sigma_{\H x c}^\sr[G_0](1,2),
\label{Sigmasrl}
\end{eqnarray}
where $\Sigma_{\H x c}^\sr[G](1,2)=\delta(1,2) v_{\H x c}^\sr[n](\b{r}_{1})$ is the local short-range Hxc self-energy.

The long-range four-point polarization propagator $\chi_{\l}^{\lr}(1,2;1',2')$ along the adiabatic connection is given by the solution of the following Bethe-Salpeter-type equation which can be derived from the Dyson equation~(\ref{Glrlm1}) by considering variations with respect to $G_\l^\lr$ [see Appendix~\ref{app:manybody}, Eq.~(\ref{DysoneqUderiv})]
\begin{eqnarray}
\left(\chi_{\l}^{\lr} \right)^{-1}(1,2;1',2') &=& \left(\chi_{\IP,\l}^\lr \right)^{-1}(1,2;1',2') 
\nonumber\\
&&- \l f_{\H x}^\lr(1,2;1',2') 
\nonumber\\
&&- f_{c,\l}^\lr(1,2;1',2'),
\label{BetheSalpetereq}
\end{eqnarray}
where $\chi_{\IP,\l}^\lr(1,2;1',2')=-i G^\lr_\l(1,2') G^\lr_\l(2,1')$ is an independent-particle (IP) polarization propagator, and $\l f_{\H x}^\lr(1,2;1',2')\! =\! i \l \, \delta \Sigma_{\H x}^{\lr}[G_\l^\lr](1,1')/\delta G_\l^\lr(2',2)$ and $f_{c,\l}^\lr(1,2;1',2') = i \, \delta \Sigma_{c,\l}^{\lr}[G_\l^\lr](1,1')/\delta G_\l^\lr(2',2)$ are long-range Hartree-exchange and correlation kernels. Note that these kernels only stem from the self-energy term $\Sigma_{\H x c,\l}^\lr[G^{\lr}_\l]$ in Eq.~(\ref{Glrlm1}) that corresponds to the two-electron interaction $\l \hat{W}_{ee}^\lr$, the other self-energy contributions which come from the one-electron terms are absorbed in the definition of $\chi_{\l}^{\lr}(1,2;1',2')$. The Hartree kernel is obtained from Eq.~(\ref{SigmaH})
\begin{eqnarray}
f_{\H}^{\lr}(1,2;1',2') =  w_{ee}^{\lr}(1,2) \delta(1,1') \delta(2,2') 
\nonumber\\
= w_{ee}^{\lr}(r_{12}) \delta(t_1-t_2) \delta(1,1') \delta(2,2'),
\label{fHlr}
\end{eqnarray}
while the HF-like exchange kernel is obtained from Eq.~(\ref{Sigmax})
\begin{eqnarray}
f_{x}^{\lr}(1,2;1',2') =  -w_{ee}^{\lr}(1,2) \delta(1,2') \delta(1',2) 
\nonumber\\
= -w_{ee}^{\lr}(r_{12}) \delta(t_1-t_2) \delta(1,2') \delta(1',2).
\label{fxlr}
\end{eqnarray}

The fluctuation-dissipation theorem is then used to express $P_{c,\l}^{\lr}$ as [see Appendix~\ref{app:manybody}, Eq.~(\ref{Pcl})]
\begin{eqnarray}
P_{c,\l}^\lr(\b{x}_1,\b{x}_2;\b{x}_1',\b{x}_2') = - \int_{-\infty}^{\infty} \frac{d\omega}{2\pi i} e^{i\omega 0^+} \Bigl[ \chi_\l^\lr(\b{x}_1,\b{x}_2;\b{x}_1',\b{x}_2';\omega)
\nonumber\\
-\chi_0(\b{x}_1,\b{x}_2;\b{x}_1',\b{x}_2';\omega) \Bigl] +\Delta_\l^\lr(\b{x}_1,\b{x}_2;\b{x}_1',\b{x}_2'), \;\;\;\;\;\;\;
\end{eqnarray}
where $\chi_\l^\lr(\b{x}_1,\b{x}_2;\b{x}_1',\b{x}_2';\omega)$ is the frequency-dependent Fourier transform of the one-time-interval polarization propagator $\chi_\l^\lr(\b{x}_1,\b{x}_2;\b{x}_1',\b{x}_2';\tau=t_1-t_2) = \chi_\l^\lr(\b{x}_1 t_1,\b{x}_2 t_2;\b{x}_1' t_1^+,\b{x}_2't_2^+)$, $\chi_0(\b{x}_1,\b{x}_2;\b{x}_1',\b{x}_2';\omega)$ is the equivalent quantity for the RSH reference Hamiltonian (at $\l=0$), and $\Delta^\lr_{\l}(\b{x}_1,\b{x}_2;\b{x}_1',\b{x}_2')$ is the contribution coming from the variation of the one-particle density matrix along the adiabatic connection. The expression of $\Delta^\lr_{\l}$ in terms of the Green functions $G^\lr_\l$ and $G_0$ is straightforward but it is sufficient to write it as $\Delta^\lr_{\l}= \Gamma[G^\lr_\l] - \Gamma[G_0]$ where $\Gamma$ is a known functional given in Appendix~\ref{app:manybody} [Eq.~(\ref{GammaG})].

So far, the theory is in principle \textit{exact}. In the following we consider two possible approximations. 
The RPA approximation
\begin{equation}
\Sigma_{xc,\l}^\lr\!=\!0,
\label{ApproximationRPA}
\end{equation}
corresponds to neglecting long-range exchange-correlation in all one-electron properties. Indeed, with this approximation, one can check that $G_{\l}^{\lr}=G_{0}$ is a solution of the Dyson equation~(\ref{Glrlm1}), i.e. the Green function remains unchanged along the adiabatic connection. It follows that $\Delta^\lr_{\l}\!=\!0$, $f_{xc,\l}^\lr\!=\!0$ and $\chi_{\IP,\l}^\lr(1,2;1',2')=-i G_0(1,2') G_0(2,1') = \chi_{0}(1,2;1',2')$. Similarly, the RPAx approximation
\begin{equation}
\Sigma_{c,\l}^\lr\!=\!0,
\label{ApproximationRPAx}
\end{equation}
corresponds to neglecting long-range correlation only in all one-electron properties. Again, this approximation implies that the Green function remains unchanged along the adiabatic connection, i.e. $G_{\l}^{\lr}=G_{0}$ and it follows that $\Delta^\lr_{\l}\!=\!0$, $f_{c,\l}^\lr\!=\!0$ and $\chi_{\IP,\l}^\lr = \chi_{0}$. As different terminologies are used in the quantum chemistry and condensed-matter physics literature, let us stress that what we call RPA here corresponds to a response equation~(\ref{BetheSalpetereq}) with no exchange-correlation kernel (and it is also sometimes called linear response time-dependent Hartree theory or \textit{direct} RPA), and what we call RPAx corresponds to a response equation with an additional HF-like exchange kernel (and it is also sometimes called linear response time-dependent Hartree-Fock theory or \textit{full} RPA).

\subsection{Expressions in an orbital basis}

The RPA or RPAx equations in an orbital basis are derived in details in Appendix~\ref{app:rpabasis}. In the basis of RSH spin orbitals, the long-range RPA or RPAx correlation energy writes
\begin{eqnarray} 
E_c^{\lr} &=& \frac{1}{2} \int_{0}^{1} d\l \sum_{ia,jb} \bra{i b} \hat{w}_{ee}^{\lr} \ket{a j} (\b{P}_{c,\l}^\lr)_{ia,jb},
\label{Eclrspinorbitals}
\end{eqnarray}
where $i$ and $j$ refer to occupied spin orbitals, and $a$ and $b$ to virtual spin orbitals, $\bra{ib} \hat{w}_{ee}^{\lr}\ket{aj}$ are the two-electron integrals with long-range interaction, and $(\b{P}^{\lr}_{c,\l})_{ia,jb}$ are the matrix elements of the correlation two-particle density matrix. The one-electron terms $v_\H^\lr$ and $v_x^\lr$ in the perturbation operator in Eq.~(\ref{Eclr}) do not contribute to $E_{c}^{\lr}$ because of the occupied-virtual/occupied-virtual structure of the two-particle density matrix in RPA or RPAx. Following the technique proposed by Furche~\cite{Fur-PRB-01}, $\b{P}^{\lr}_{c,\l}$ can be obtained as
\begin{eqnarray} 
\b{P}_{c,\l}^\lr = \left( \b{A}^\lr_\l-\b{B}^\lr_\l \right)^{1/2} \left( \b{M}^\lr_\l \right)^{-1/2} \left( \b{A}^\lr_\l-\b{B}^\lr_\l \right)^{1/2} -\b{1}.
\label{Pcllr}
\end{eqnarray}
with $\b{M}^\lr_{\l}=\left( \b{A}^\lr_\l-\b{B}^\lr_\l \right)^{1/2} \left( \b{A}^\lr_\l+\b{B}^\lr_\l \right) \left( \b{A}^\lr_\l-\b{B}^\lr_\l \right)^{1/2}$, and the orbital rotation Hessians
\begin{subequations}
\begin{eqnarray}
\left( \b{A}^\lr_\l \right)_{ia,jb} &=&  (\epsilon_a-\epsilon_i) \delta_{ij} \delta_{ab} 
\nonumber\\
&&+ \l \left [ \bra{i b} \hat{w}_{ee}^\lr \ket{a j} - \xi \bra{i b} \hat{w}_{ee}^\lr \ket{j a} \right],
\label{}
\end{eqnarray}
\begin{eqnarray}
\left( \b{B}^\lr_\l \right)_{ia,jb} =  \l \left [ \bra{a b} \hat{w}_{ee}^\lr \ket{i j} - \xi \bra{a b} \hat{w}_{ee}^\lr \ket{j i} \right].
\label{}
\end{eqnarray}
\label{Hessians}
\end{subequations}
where $\epsilon_{i}$ are the RSH orbital eigenvalues, and $\xi=0$ or $\xi=1$ for RPA and RPAx, respectively. For spin-restricted closed-shell calculations, the correlation energy writes in terms of spatial orbitals
\begin{eqnarray} 
E_c^\lr &=& \frac{1}{2} \int_{0}^{1} d\l \sum_{ia,jb} \bra{i b} \hat{w}_{ee}^\lr \ket{a j} (^1\b{P}_{c,\l}^\lr)_{ia,jb},
\label{Eclrorbitals}
\end{eqnarray}
where $i$ and $j$ now refer to occupied spatial orbitals, and $a$ and $b$ to virtual spatial orbitals, and $^1\b{P}_{c,\l}^\lr$ is the spin-singlet-adapted correlation two-particle density matrix obtained as
\begin{eqnarray} 
^1\b{P}_{c,\l}^\lr = 2 \left[ \left( {^1}\b{A}^\lr_\l-{^1}\b{B}^\lr_\l \right)^{1/2} \left( {^1}\b{M}^\lr_\l \right)^{-1/2} \left( {^1}\b{A}^\lr_\l-{^1}\b{B}^\lr_\l \right)^{1/2} -\b{1} \right], \;\;
\label{1Pcllr}
\end{eqnarray}
with ${^1}\b{M}^\lr_{\l}=\left( {^1}\b{A}^\lr_\l-{^1}\b{B}^\lr_\l \right)^{1/2} \left( {^1}\b{A}^\lr_\l+{^1}\b{B}^\lr_\l \right) \left( {^1}\b{A}^\lr_\l-{^1}\b{B}^\lr_\l \right)^{1/2}$, and the singlet orbital rotation Hessians
\begin{subequations}
\begin{eqnarray}
\left( ^1\b{A}^\lr_\l \right)_{ia,jb} &=&  (\epsilon_a-\epsilon_i) \delta_{ij} \delta_{ab} 
\nonumber\\
&&+ \l \left [ 2 \bra{i b} \hat{w}_{ee}^\lr \ket{a j} - \xi \bra{i b} \hat{w}_{ee}^\lr \ket{j a} \right],
\label{}
\end{eqnarray}
\begin{eqnarray}
\left( ^1\b{B}^\lr_\l \right)_{ia,jb} =  \l \left [ 2 \bra{a b} \hat{w}_{ee}^\lr \ket{i j} - \xi \bra{a b} \hat{w}_{ee}^\lr \ket{j i} \right].
\end{eqnarray}
\label{Hessianssinglet}
\end{subequations}
Only singlet excitations contribute to Eq.~(\ref{Eclrorbitals}), since the two-electron integrals involved vanish for triplet excitations. 

In Eq.~(\ref{Pcllr}), it is assumed that $\b{A}^\lr_\l+\b{B}^\lr_\l$ and $\b{A}^\lr_\l-\b{B}^\lr_\l$ are positive definite. In RPA, this is always the case. On the contrary, in RPAx, this is not always the case, i.e. instabilities can be encountered, and Eq.~(\ref{Pcllr}) can fail. In spin-restricted closed-shell formalism, one may encounter singlet instabilities in the RPAx theory defined here, for example when dissociating a bond, but not triplet instabilities since triplet excitations do not contribute at all.
In practice, singlet instabilities are usually not encountered for weakly-interacting closed-shell systems. Note that other variants of RPA-type correlation energy expressions using a HF exchange response kernel, such as the plasmon formula~\cite{MclBal-RMP-64,SzaOst-JCP-77,ScuHenSor-JCP-08} or the equivalent ring coupled-cluster-doubles theory~\cite{ScuHenSor-JCP-08}, require contributions from both singlet and triplet excitations, and are thus subject to triplet instabilities (e.g. in a system such as Be$_2$).

Similarly to the notation used in Ref.~\onlinecite{TouGerJanSavAng-PRL-09}, the range-separated method obtained by adding to the RSH energy the long-range RPAx correlation energy [$\xi=1$ in Eqs.~(\ref{Hessians}) or Eqs.~(\ref{Hessianssinglet}] will be referred to as RSH+lrRPAx. For consistency, the range-separated method obtained by adding to the RSH energy the long-range RPA correlation energy [$\xi=0$ in Eqs.~(\ref{Hessians}) or Eqs.~(\ref{Hessianssinglet})] will be referred to as RSH+lrRPA, although it is equivalent to the method called ``LC-$\omega$LDA+dRPA'' in Refs.~\onlinecite{JanHenScu-JCP-09,JanHenScu-JCP-09b,JanScu-JCP-09,PaiJanHenScuGruKre-JCP-10} in the special case of the short-range LDA functional. At second order in the electron-electron interaction, the RSH+lrRPAx method reduces to the range-separated method of Ref.~\onlinecite{AngGerSavTou-PRA-05} based on long-range second-order M{\o}ller-Plesset perturbation theory, to which we will refer as RSH+lrMP2. Since RPA approaches can be seen as simple approximations to coupled-cluster theory~\cite{ScuHenSor-JCP-08}, the RSH+lrRPA and RSH+lrRPAx methods bear some resemblance to the range-separated method of Ref.~\onlinecite{GolWerSto-PCCP-05} where the long-range correlation energy is evaluated by coupled-cluster theory (with single, double and perturbative triple excitations), to which we will refer as RSH+lrCCSD(T).

We note that one can develop long-rang many-body perturbation theories starting from other references than the RSH reference. For example, starting from the usual (approximate) Kohn-Sham reference could be appropriate for solid-state systems. For the finite systems considered here, RSH is a good reference, as confirmed by other authors~\cite{JanScu-JCP-09}.

\section{Computational details}

All calculations have been performed with a development version of MOLPRO 2008~\cite{Molproshort-PROG-08}, implementing equations~(\ref{Eclrorbitals})-(\ref{Hessianssinglet}). We first perform a self-consistent RSH calculation with the short-range PBE xc functional of Ref.~\onlinecite{GolWerSto-PCCP-05} (this RSH calculation could also be referred to as ``lrHF+srPBE'', a notation closer to the one used by other authors~\cite{GolWerSto-PCCP-05}) and add the long-range MP2, RPA, RPAx or CCSD(T) correlation energy calculated with RSH orbitals. For RPA or RPAx, the $\l$-integration in Eq.~(\ref{Eclrorbitals}) is done by a 7-point Gauss-Legendre quadrature~\cite{Fur-PRB-01}. The range separation parameter is taken at $\mu=0.5$ bohr$^{-1}$, in agreement with previous studies~\cite{GerAng-CPL-05a}, without trying to adjust it for each system. To show the dependence on the orbitals, the full-range RPA calculations have been done with PBE~\cite{PerBurErn-PRL-96} and HF orbitals, which will be denoted by PBE+RPA and HF+RPA, respectively~\cite{TouZhuAngSav-JJJ-XX-note2}. The full-range MP2, RPAx and CCSD(T) calculations have been done with HF orbitals, and thus, for notation consistency, will be denoted by HF+MP2, HF+RPAx and HF+CCSD(T), respectively. We use large Dunning basis sets~\cite{Dun-JCP-89,WooDun-JCP-93,WooDun-JCP-94,Fel-JCC-96,WilWooPetDun-JCP-99,KopPet-JPCA-02,SchDidElsSunGurChaLiWin-JCIM-07}. Core electrons are kept frozen in all the full-range and range-separated MP2, RPA, RPAx and CCSD(T) calculations (i.e. only excitations of valence electrons are considered). The basis set superposition error (BSSE) is removed by the counterpoise method. For the alkaline-earth dimers, it has been checked than adding diffuse basis functions or core excitations do not change significantly the results. Extrapolations to the complete basis set (CBS) limit have also been considered for some systems. For the full-range methods, the standard three-point exponential formula for the HF (or KS) reference $E_{\HF}(n) = E_{\HF}(\CBS) + A e^{-B n}$ with the cardinal number $n=3,4,5$, and two-point formula for the correlation energy $E_{c}(n) = E_{c}(\CBS) + C/n^3$ with $n=4,5$ have been used. For the range-separated methods, we have also used these two formulas for the RSH reference and the long-range correlation energy, even though in this case the dependence on the cardinal number would deserve a detailed study.

For each dimer interaction energy curve, we choose 16 to 20 intermolecular distances, with denser sampling around the equilibrium distance. A third-order polynomial is used for interpolation. The hard core radius is taken as the distance where the interaction energy is 0, and the equilibrium distance and binding energy are from the minimum of the interpolated interaction energy curve. The harmonic vibrational frequency is obtained from the second-order derivative of the energy curve at the equilibrium distance. For $C_6$ dispersion coefficients, the interaction energy $E_\text{int}$ is calculated at seven extra distances $R_i$ from 30 to 60 bohr, and the coefficient is estimated by averaging with the following formula
\begin{eqnarray} 
C_6=\exp \left( \frac{1}{7}\sum_{i=1}^7 \left( \ln|E_{\text{int}}(R_i)|+6 \ln(R_i) \right) \right),
\end{eqnarray} 
similarly to what has been done in Ref.~\onlinecite{JanHenScu-JCP-09b}.

\begin{figure}
\includegraphics[scale=0.30,angle=-90]{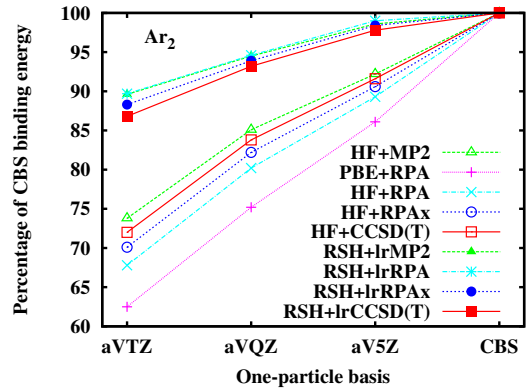}
\caption{(Color online) Basis set dependence of the equilibrium binding energy of Ar$_2$ for different full-range and range-separated methods, presented as the percentage of the binding energy recovered with respect to the CBS limit (aVTZ, aVQZ and aV5Z stand for aug-cc-pVTZ, aug-cc-pVQZ and aug-cc-pV5Z, respectively). 
}
\label{fig:basis}
\end{figure}

\begin{figure*}
\includegraphics[scale=0.30,angle=-90]{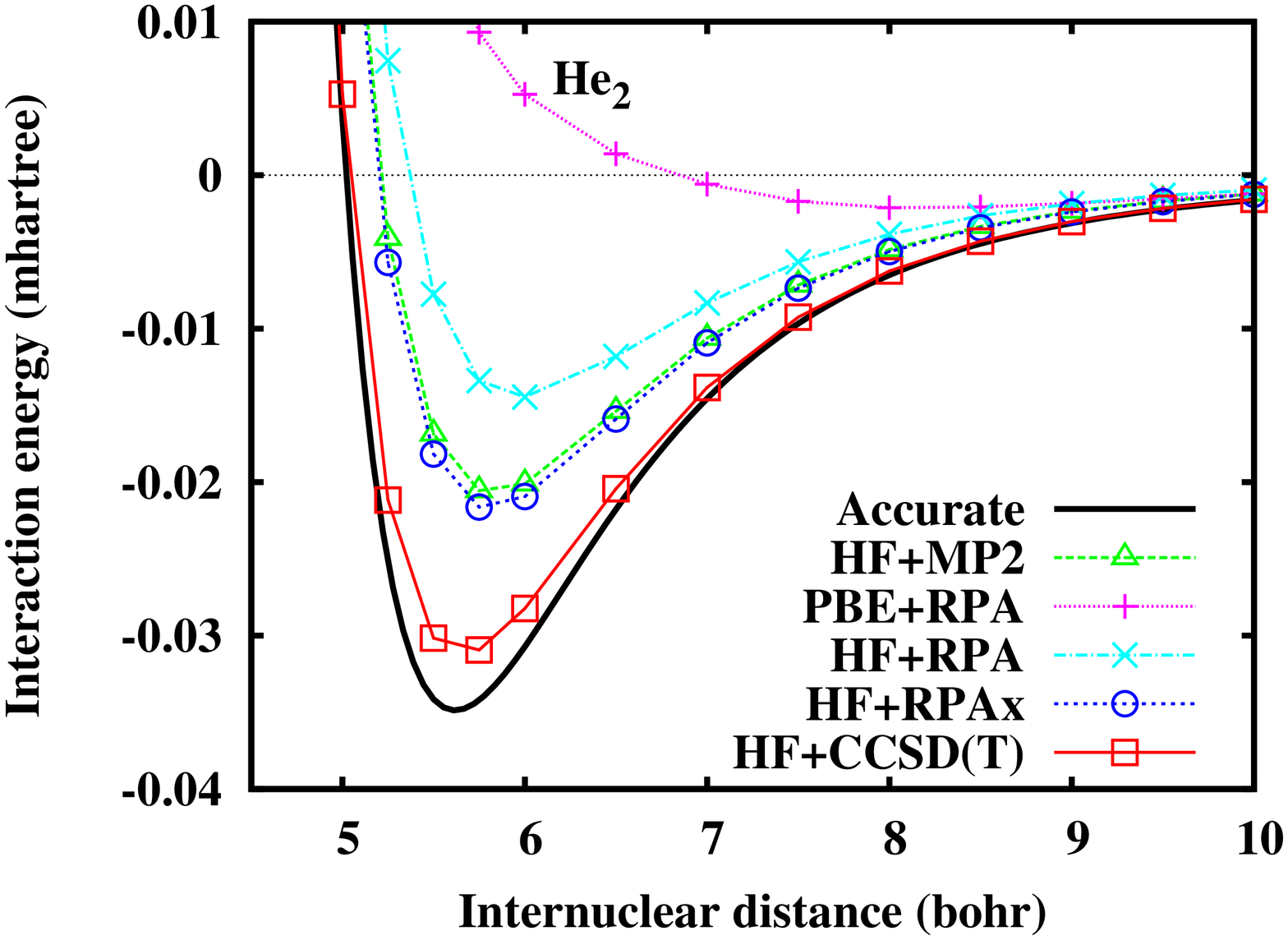}
\includegraphics[scale=0.30,angle=-90]{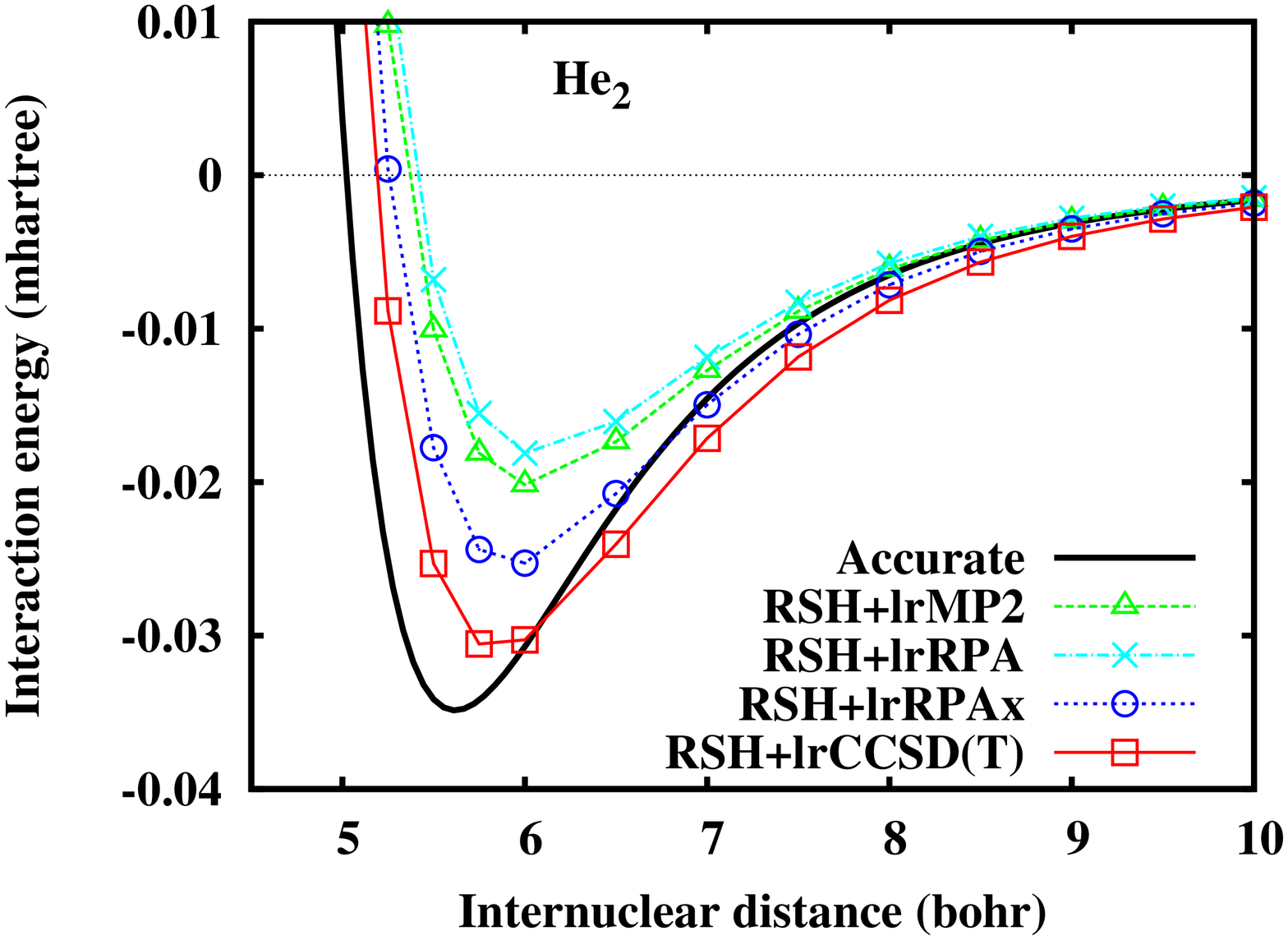}
\includegraphics[scale=0.30,angle=-90]{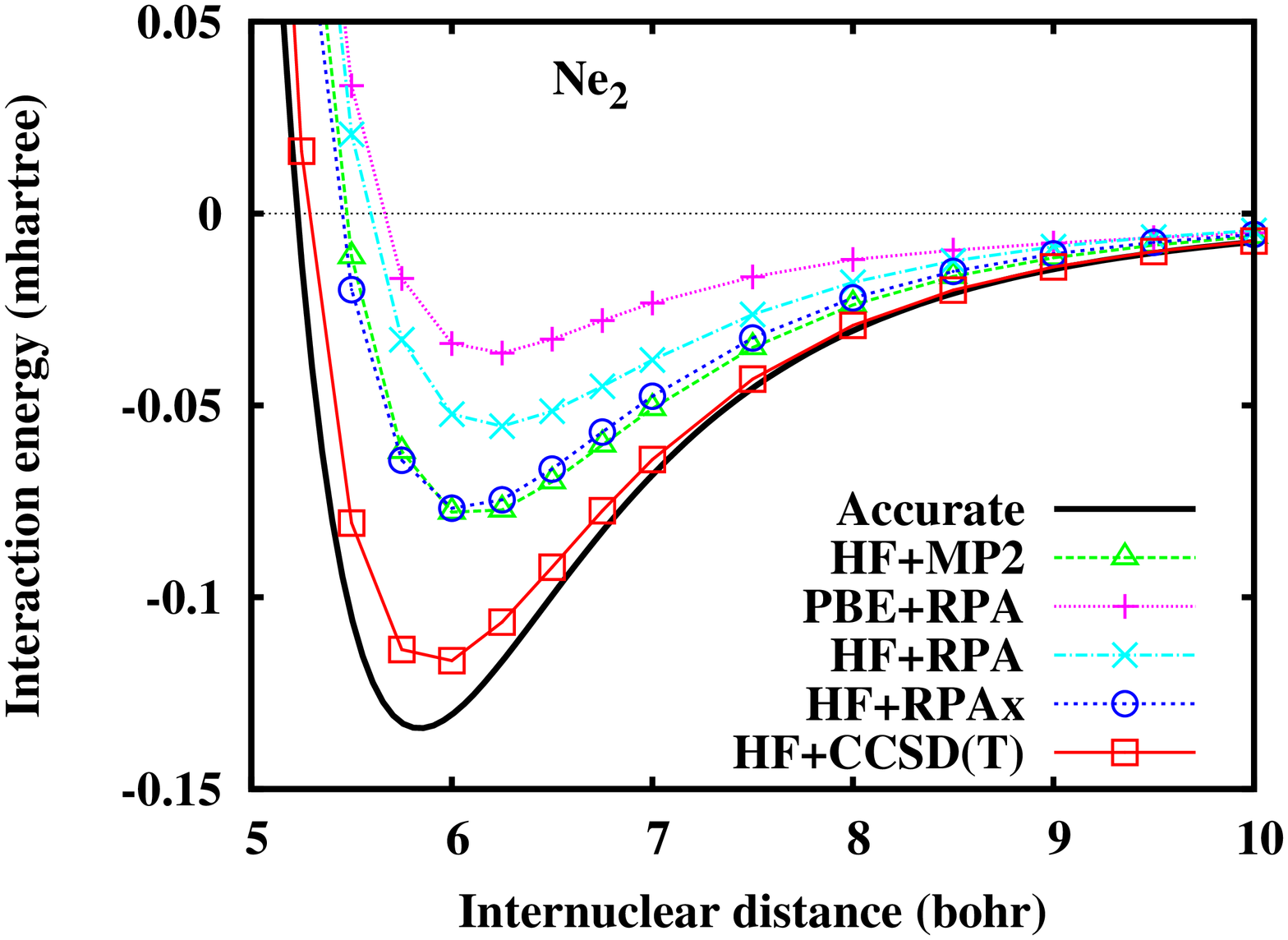}
\includegraphics[scale=0.30,angle=-90]{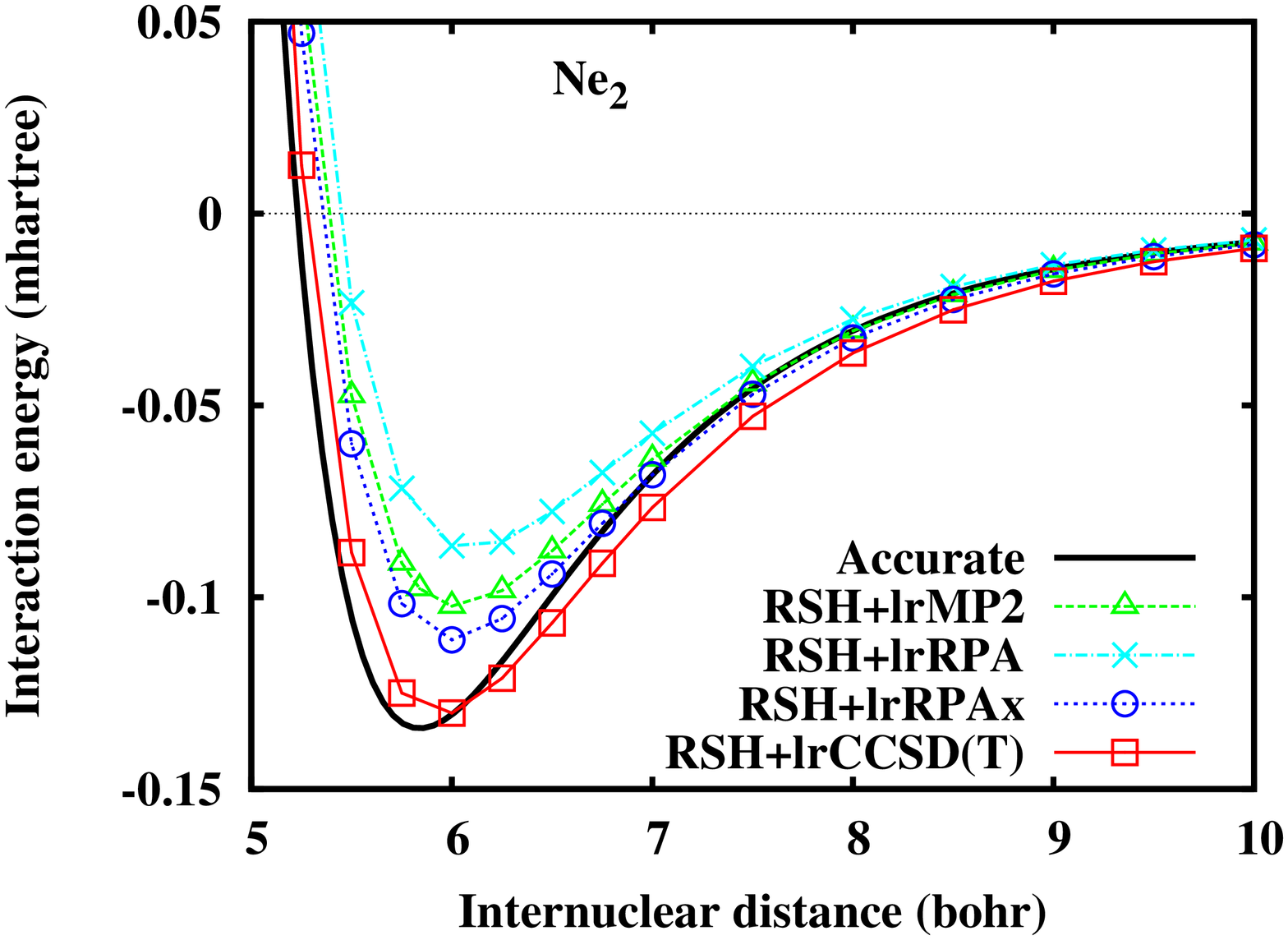}
\includegraphics[scale=0.30,angle=-90]{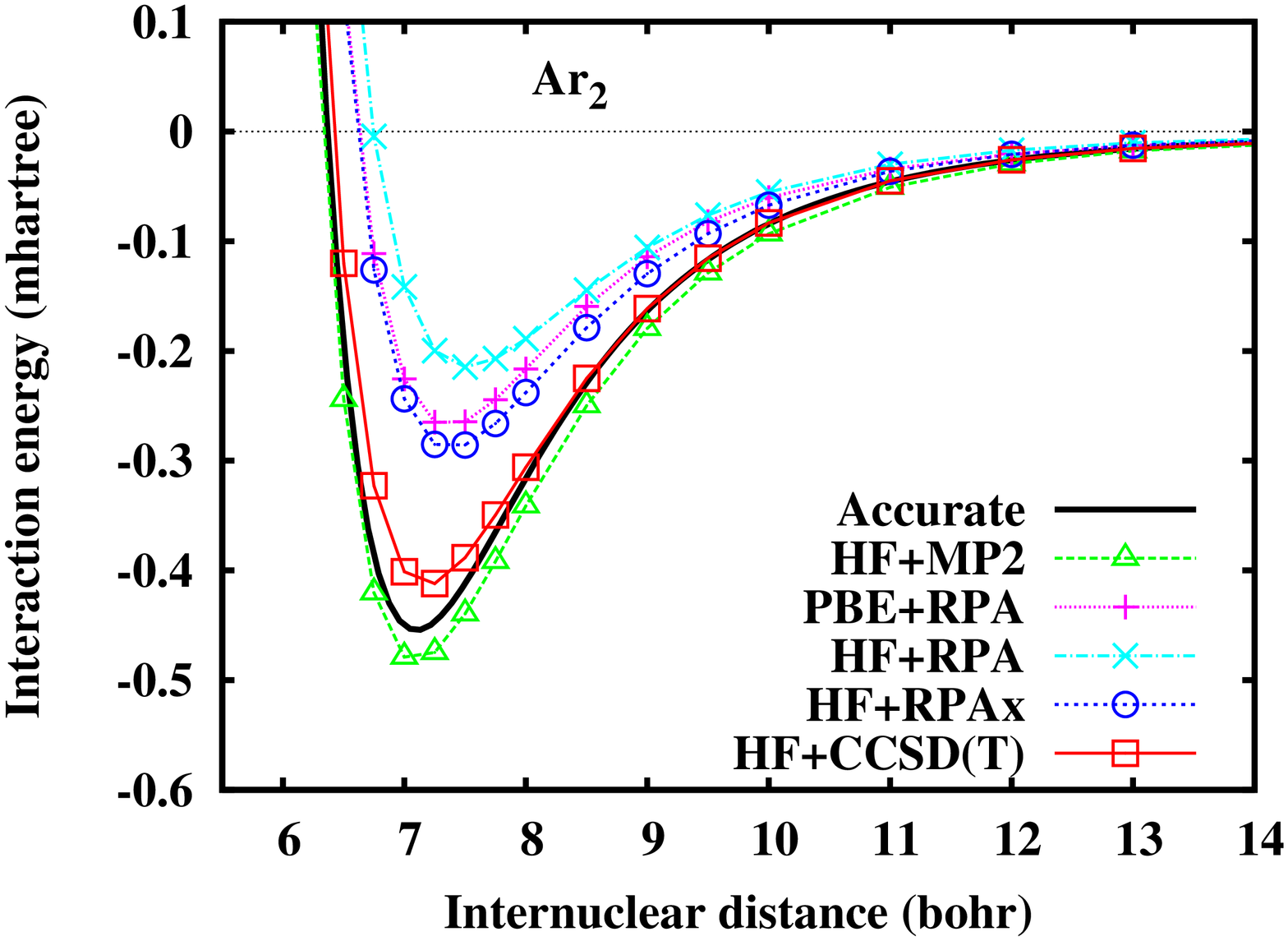}
\includegraphics[scale=0.30,angle=-90]{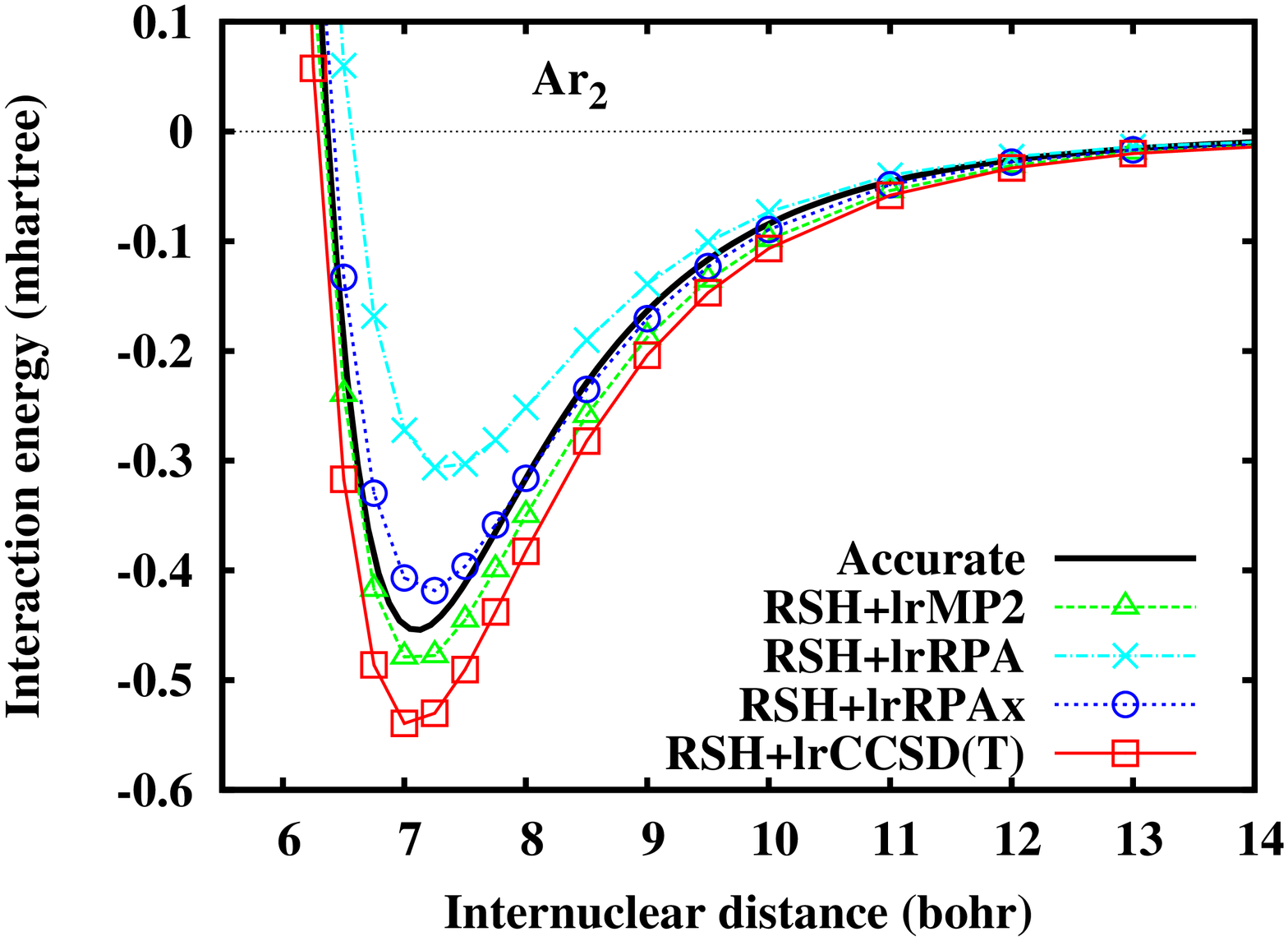}
\includegraphics[scale=0.30,angle=-90]{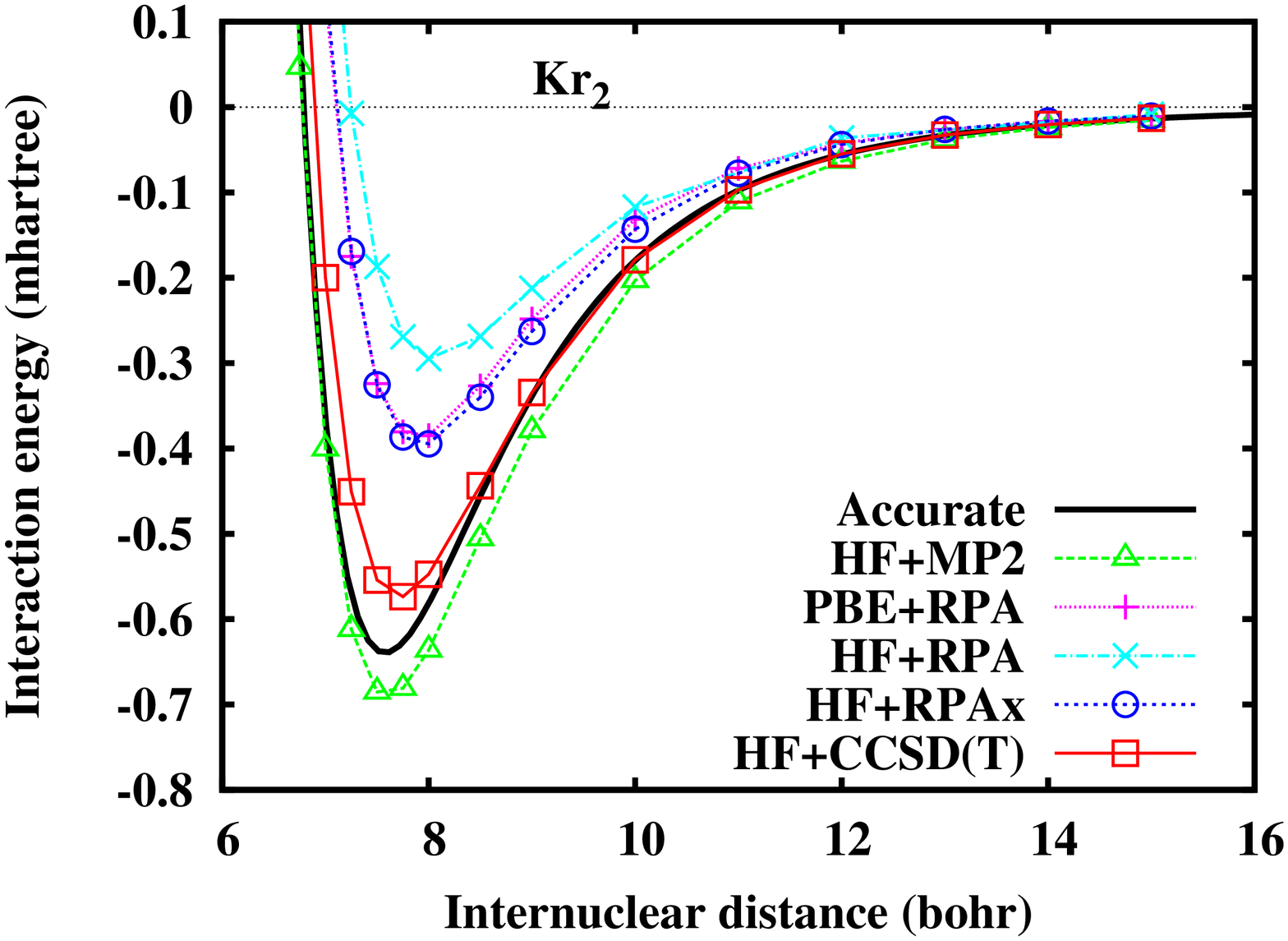}
\includegraphics[scale=0.30,angle=-90]{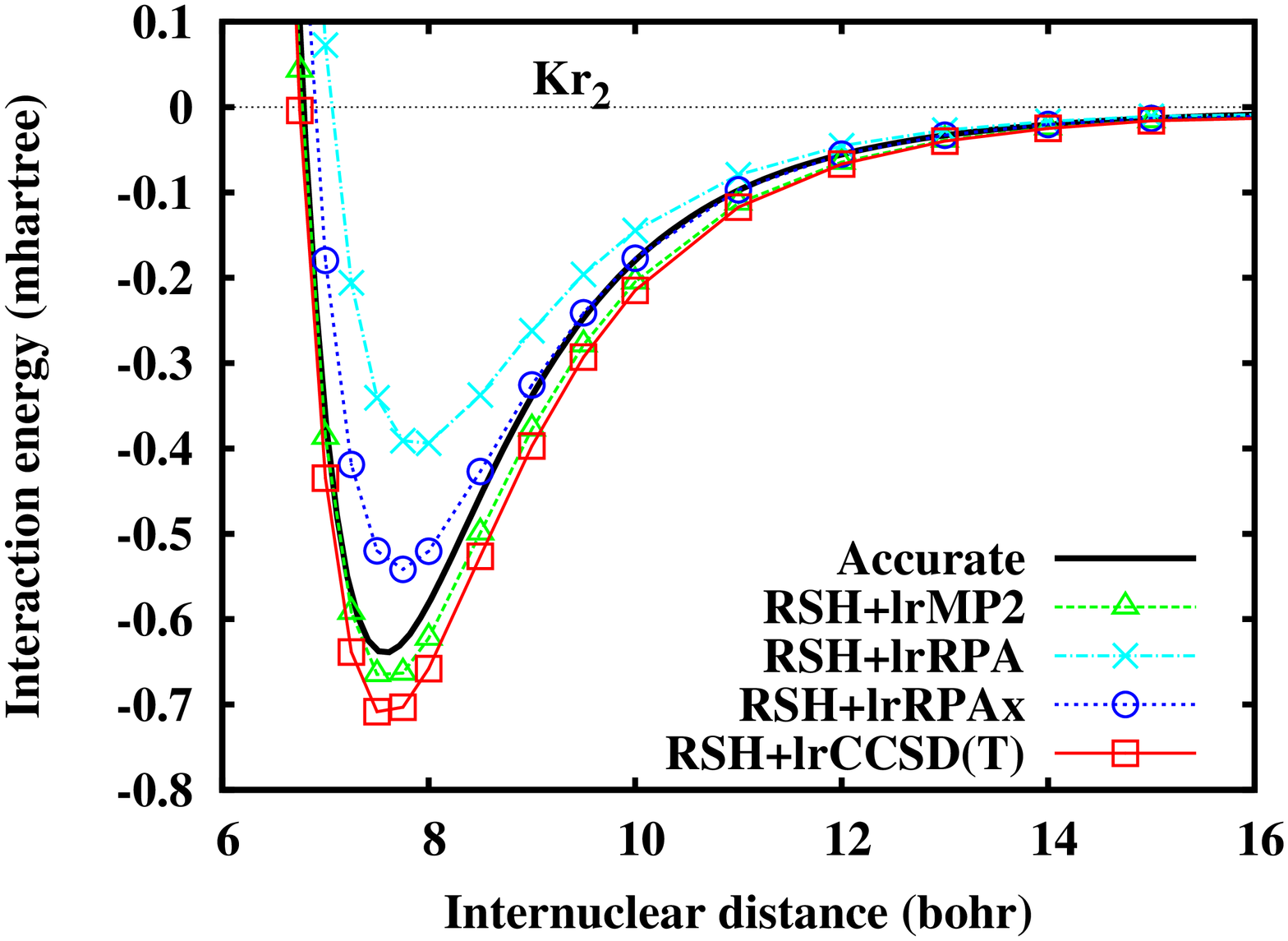}
\caption{(Color online) Interaction energy curves of He$_2$, Ne$_2$, Ar$_2$ and Kr$_2$ calculated by different full-range (left) and range-separated (right) methods.
The basis is aug-cc-pV5Z. The accurate curves are from Ref.~\onlinecite{TanToe-JCP-03}.
}
\label{fig:raregas}
\end{figure*}

\begingroup
\squeezetable
\begin{table*}[t]
\caption{Hard-core radii $\sigma$ (bohr), equilibrium distances $R_e$ (bohr), equilibrium binding energies $D_e$ (mhartree), harmonic vibrational frequencies $\omega_e$ (cm$^{-1}$) and dispersion coefficients $C_6$ for ten homonuclear and heteronuclear rare-gas dimers from different full-range and range-separated methods with aug-cc-pV5Z basis. Mean absolute percentage errors (MA\%E) are also given.}
\label{tab:raregas}
\begin{tabular}{lrrrrrrrrrr} \hline \hline
          &  HF+MP2 &  PBE+RPA& HF+RPA &    HF+RPAx  & HF+CCSD(T)& RSH+lrMP2& RSH+lrRPA & RSH+lrRPAx & RSH+lrCCSD(T) & Estimated exact$^a$ \\
\hline
\multicolumn{10}{l}{He$_2$}\\
$\sigma$  & 5.20   &  6.81   &  5.34    &  5.18   &  5.03   &  5.35   &  5.39   &  5.25   &  5.17      &    5.02  \\
$R_e$     & 5.83   &  8.16   &  5.95    &  5.82   &  5.65   &  6.00   &  6.10   &  5.92   &  5.85      &    5.62  \\
$D_e$     & 0.0208 &  0.0021 &  0.0145  &  0.0218 &  0.0313 &  0.0202 &  0.0183 &  0.0255 &  0.0309    &   0.0348 \\ 
$\omega_e$& 26.9   &   4.5   &  24.1    &  27.4   &  33.6   &  26.2   &  22.3   &  28.6   &  30.4      &   34.3   \\
$C_6$     & 1.13   &  1.36   &  0.88    &  1.14   &  1.46   &  1.42   &  1.34   &  1.67   &  1.91      &   1.461  \\
\hline
\multicolumn{10}{l}{He-Ne}\\
$\sigma$  & 5.32   &  5.81   &  5.44    &  5.29   &  5.13   &  5.33   &  5.38   &  5.27   &  5.19      &    5.16  \\
$R_e$     & 5.95   &  6.37   &  6.08    &  5.91   &  5.77   &  5.99   &  6.07   &  5.93   &  5.87      &    5.76  \\
$D_e$     & 0.0401 &  0.0064 &  0.0284  &  0.0410 &  0.0609 &  0.0458 &  0.0401 &  0.0533 &  0.0638    &   0.0660 \\ 
$\omega_e$& 28.8   &  13.0   &  23.8    &  29.5   &  34.3   &  28.4   &  26.2   &  30.9   &  33.5      &   36.1   \\
$C_6$     & 2.43   &  2.77   &  1.84    &  2.32   &  3.07   &  3.12   &  2.84   &  3.44   &  4.04      &   3.029  \\
\hline
\multicolumn{10}{l}{He-Ar}\\
$\sigma$  & 6.02   &  6.31   &  6.27    &  6.11   &  5.92   &  6.01   &  6.14   &  5.99   &  5.87      &    5.92  \\
$R_e$     & 6.73   &  6.96   &  6.97    &  6.83   &  6.64   &  6.77   &  6.89   &  6.73   &  6.63      &    6.61  \\
$D_e$     & 0.0736 &  0.0307 &  0.0424  &  0.0608 &  0.0874 &  0.0808 &  0.0616 &  0.0854 &  0.1071    &   0.0937 \\ 
$\omega_e$& 32.3   &  24.1   &  25.9    &  29.4   &  35.7   &  31.5   &  29.0   &  33.3   &  37.4      &   36.0   \\
$C_6$     & 9.1    &  9.1    &  6.1     &  7.6    & 11.6    & 10.6    &  8.7    & 10.8    & 12.6       &  9.538   \\
\hline
\multicolumn{10}{l}{He-Kr}\\
$\sigma$  & 6.38   &  6.67   &  6.67    &  6.50   &  6.28   &  6.35   &  6.52   &  6.34   &  6.22      &    6.25  \\
$R_e$     & 7.15   &  7.37   &  7.42    &  7.26   &  7.05   &  7.14   &  7.31   &  7.13   &  7.03      &    6.98  \\
$D_e$     & 0.0747 &  0.0337 &  0.0423  &  0.0606 &  0.0881 &  0.0833 &  0.0613 &  0.0857 &  0.1084    &   0.0996 \\ 
$\omega_e$& 30.1   &  22.3   &  23.4    &  26.3   &  32.4   &  30.7   &  25.9   &  31.2   &  34.2      &   33.7   \\
$C_6$     & 12.9   &  12.5   &  8.5     &  10.7   &  14.0   &  14.9   &  12.0   &  14.7   &  17.3      &   13.40  \\
\hline
\multicolumn{10}{l}{Ne$_2$}\\
$\sigma$  & 5.47   &  5.63   &  5.57    &  5.43   &  5.28   &  5.36   &  5.43   &  5.33   &  5.27      &    5.23  \\
$R_e$     & 6.11   &  6.18   &  6.19    &  6.07   &  5.90   &  6.03   &  6.10   &  5.98   &  5.93      &    5.84  \\
$D_e$     & 0.079  &  0.037  &  0.056   &  0.077  &  0.118  &  0.102  &  0.088  &  0.111  &  0.131     &   0.134  \\ 
$\omega_e$& 22.8   &  18.7   &  19.7    &  22.6   &  28.8   &  23.8   &  22.9   &  25.9   &  28.3      &   29.4   \\
$C_6$     & 5.24   &  6.84   &  3.91    &  4.77   &  6.35   &  6.80   &  6.10   &  7.03   &  8.08      &   6.383  \\
\hline
\multicolumn{10}{l}{Ne-Ar}\\
$\sigma$  & 6.02   &  6.21   &  6.28    &  6.13   &  5.94   &  5.92   &  6.06   &  5.93   &  5.84      &    5.89  \\
$R_e$     & 6.72   &  6.87   &  7.01    &  6.85   &  6.65   &  6.66   &  6.80   &  6.67   &  6.59      &    6.57  \\
$D_e$     & 0.163  &  0.095  &  0.092   &  0.126  &  0.189  &  0.196  &  0.147  &  0.192  &  0.235     &   0.211  \\ 
$\omega_e$& 25.3   &  21.6   &  17.4    &  22.6   &  27.7   &  27.2   &  23.0   &  26.9   &  29.3      &   28.7   \\
$C_6$     & 19.2   &  18.9   &  12.5    &  15.2   &  18.2   &  22.6   &  18.3   &  21.8   &  25.3      &   19.50  \\
\hline
\multicolumn{10}{l}{Ne-Kr}\\
$\sigma$  & 6.31   &  6.53   &  6.61    &  6.46   &  6.24   &  6.20   &  6.36   &  6.23   &  6.14      &    6.17  \\
$R_e$     & 7.08   &  7.21   &  7.36    &  7.20   &  6.98   &  6.97   &  7.13   &  7.01   &  6.91      &    6.89  \\
$D_e$     & 0.174  &  0.104  &  0.096   &  0.131  &  0.201  &  0.212  &  0.153  &  0.201  &  0.248     &   0.224  \\ 
$\omega_e$& 22.4   &  19.0   &  17.0    &  19.8   &  24.5   &  24.4   &  20.7   &  23.1   &  26.5      &   25.3   \\
$C_6$     & 27.0   &  26.2   &  17.4    &  21.1   &  27.4   &  31.5   &  24.8   &  29.5   &  34.0      &   27.30  \\
\hline
\multicolumn{10}{l}{Ar$_2$}\\
$\sigma$  & 6.32   &  6.61   &  6.74    &  6.60   &  6.41   &  6.32   &  6.55   &  6.40   &  6.28      &    6.37  \\
$R_e$     & 7.10   &  7.36   &  7.52    &  7.37   &  7.17   &  7.11   &  7.34   &  7.18   &  7.07      &    7.10  \\
$D_e$     & 0.483  &  0.269  &  0.215   &  0.289  &  0.414  &  0.484  &  0.308  &  0.420  &  0.542     &   0.454  \\ 
$\omega_e$& 32.7   &  25.5   &  21.4    &  25.5   &  30.7   &  32.1   &  25.5   &  30.0   &  33.5      &   32.1   \\
$C_6$     & 76.3   &  58.6   &  42.9    &  52.0   &  64.5   &  80.7   &  57.4   &  69.6   &  85.0      &   64.30  \\
\hline
\multicolumn{10}{l}{Ar-Kr}\\
$\sigma$  & 6.55   &  6.85   &  7.00    &  6.85   &  6.65   &  6.55   &  6.80   &  6.64   &  6.52      &    6.59  \\
$R_e$     & 7.36   &  7.64   &  7.81    &  7.66   &  7.45   &  7.37   &  7.62   &  7.46   &  7.34      &    7.35  \\
$D_e$     & 0.570  &  0.319  &  0.248   &  0.334  &  0.481  &  0.563  &  0.346  &  0.472  &  0.615     &   0.531  \\ 
$\omega_e$& 29.5   &  22.9   &  19.4    &  22.7   &  27.3   &  28.7   &  22.5   &  26.3   &  29.8      &   28.6   \\
$C_6$     & 109.9  &  82.1   &  60.7    &  73.6   &  94.8   & 114.1   &  80.0   &  97.4   & 117.1      &   91.13  \\
\hline
\multicolumn{10}{l}{Kr$_2$}\\
$\sigma$  & 6.77   &  7.09   &  7.24    &  7.10   &  6.88   &  6.77   &  7.05   &  6.88   &  6.75      &    6.79  \\
$R_e$     & 7.60   &  7.90   &  8.08    &  7.92   &  7.70   &  7.61   &  7.89   &  7.72   &  7.60      &    7.58  \\
$D_e$     & 0.691  &  0.388  &  0.296   &  0.396  &  0.575  &  0.671  &  0.397  &  0.542  &  0.713     &   0.638  \\ 
$\omega_e$& 25.1   &  19.8   &  16.2    &  19.7   &  23.2   &  24.4   &  19.2   &  21.9   &  25.0      &   24.4   \\
$C_6$     & 159    &  116    &  86      &  105    &  132    &  162    &  109    &  134    &  163       &   129.6  \\
\hline
\multicolumn{10}{l}{MA\%E (\%)}\\
$\sigma$  & 2.1    &  9.3    &  6.3     &  3.8    &  0.7    &  1.8    &  3.9    &  1.5    &  1.0       &   0.0    \\
$R_e$     & 2.1    &  9.4    &  6.1     &  3.9    &  1.0    &  2.2    &  4.5    &  2.2    &  1.0       &   0.0    \\
$D_e$     & 23     &  62     &  56      &  39     &  10     &  16     &  36     &  14     &  11        &   0.0    \\
$\omega_e$& 12     &  36     &  33      &  21     &  3.4    &  9.5    &  23     &  10     &  4.6       &   0.0    \\
$C_6$     & 13     &  7.0    &  36      &  22     &  4.1    &  14     &  9.2    &  10     &  29        &   0.0    \\  
\hline\hline
$^a$ From Ref.~\onlinecite{TanToe-JCP-03}\\
\end{tabular}
\end{table*}
\endgroup

\section{Applications}

\subsection{Basis set dependence}

The convergence of the equilibrium binding energy of Ar$_2$ with respect to the basis set size up to the CBS limit for the full-range methods HF+MP2, PBE+RPA, HF+RPA, HF+CCSD(T) and for the range-separated methods RSH+lrMP2, RSH+lrRPA, RSH+lrRPAx, RSH+lrCCSD(T) is represented in Fig.~\ref{fig:basis}. Full-range RPA with PBE orbitals has a very strong dependence on the basis size, as already noted (e.g. Refs.~\onlinecite{Fur-PRB-01,TouGerJanSavAng-PRL-09}). Full-range RPA with HF orbitals has a bit weaker basis dependence, similar to full-range HF+MP2, HF+RPAx and HF+CCSD(T). All the range-separated methods have essentially identical, very favorable basis set convergence. Since the slow convergence of full-range methods is related to the explicit description of short-range correlation, it is not surprising that range-separated methods have a faster convergence because they leave the description of short-range correlation to the short-range density functional. These results are consistent with other studies, e.g. Refs.~\onlinecite{JanHenScu-JCP-09b,PaiJanHenScuGruKre-JCP-10}. Note that, with the aug-cc-pV5Z basis set, all the range-separated methods are essentially converged (98-99\% of the CBS binding energy), therefore we will not use CBS extrapolations in the following. However, one should keep in mind that with this basis set the full-range methods are not yet fully converged, with about 90\% of the CBS binding energy.

\begin{figure*}
\includegraphics[scale=0.30,angle=-90]{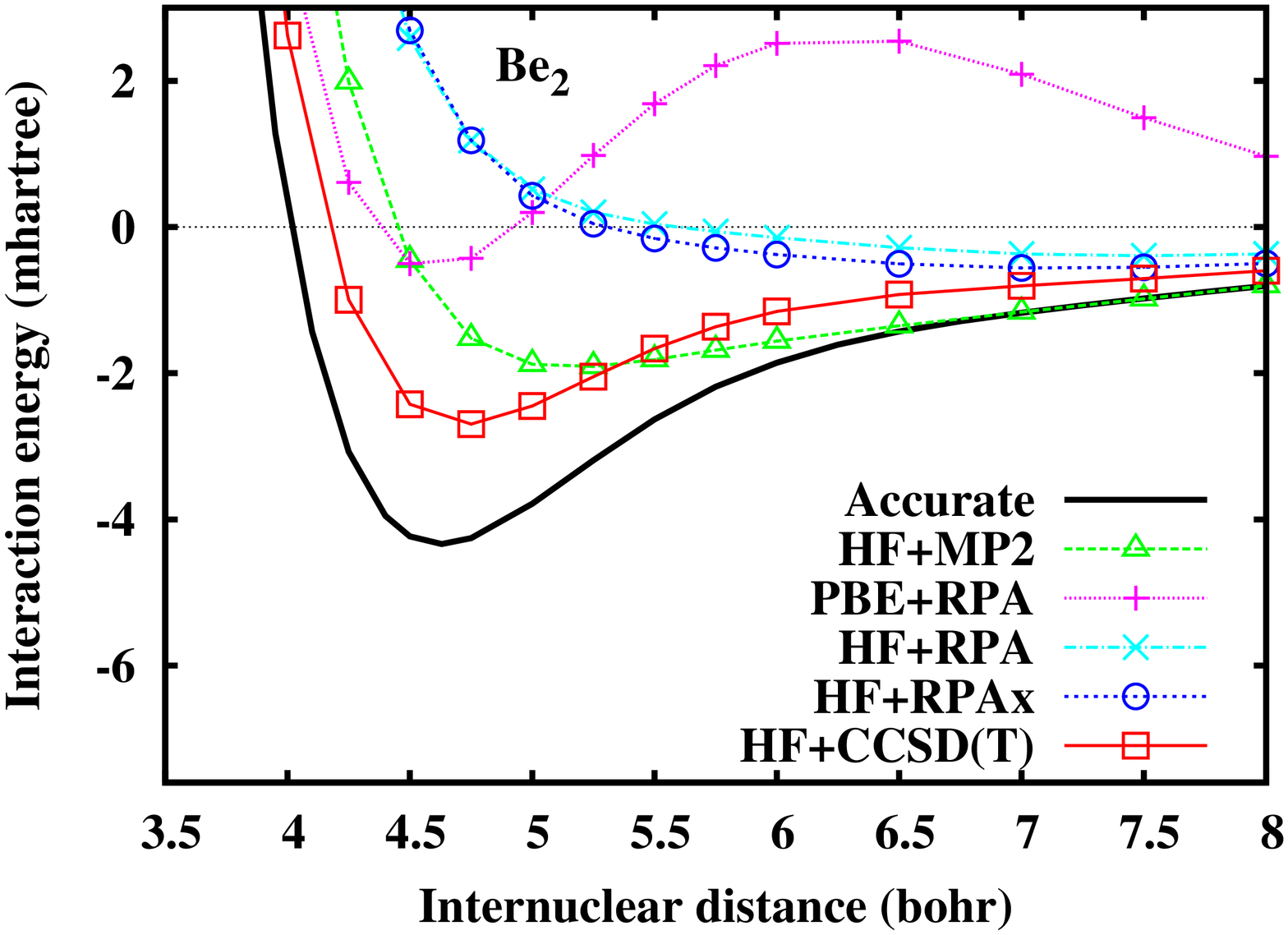}
\includegraphics[scale=0.30,angle=-90]{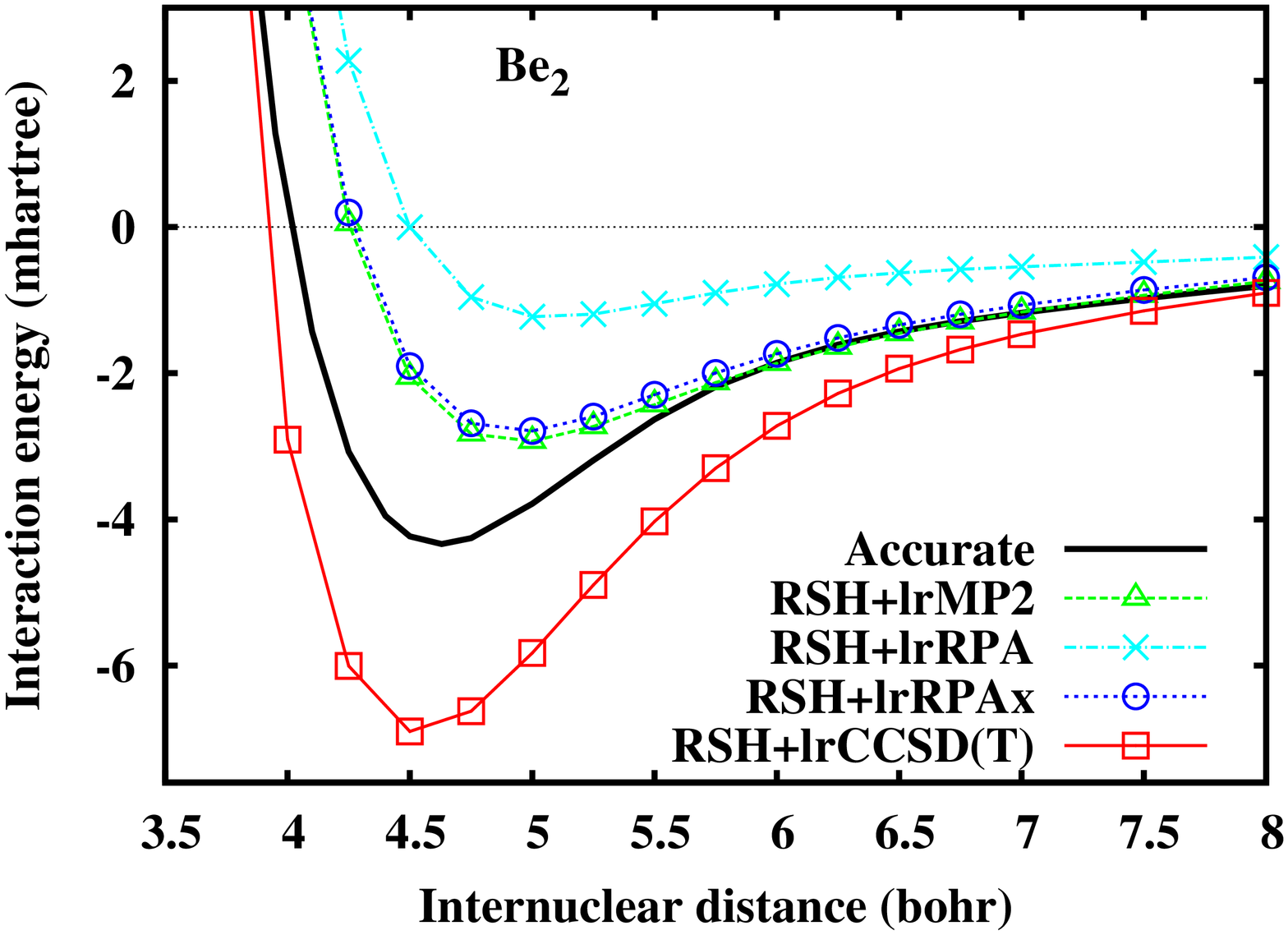}
\includegraphics[scale=0.30,angle=-90]{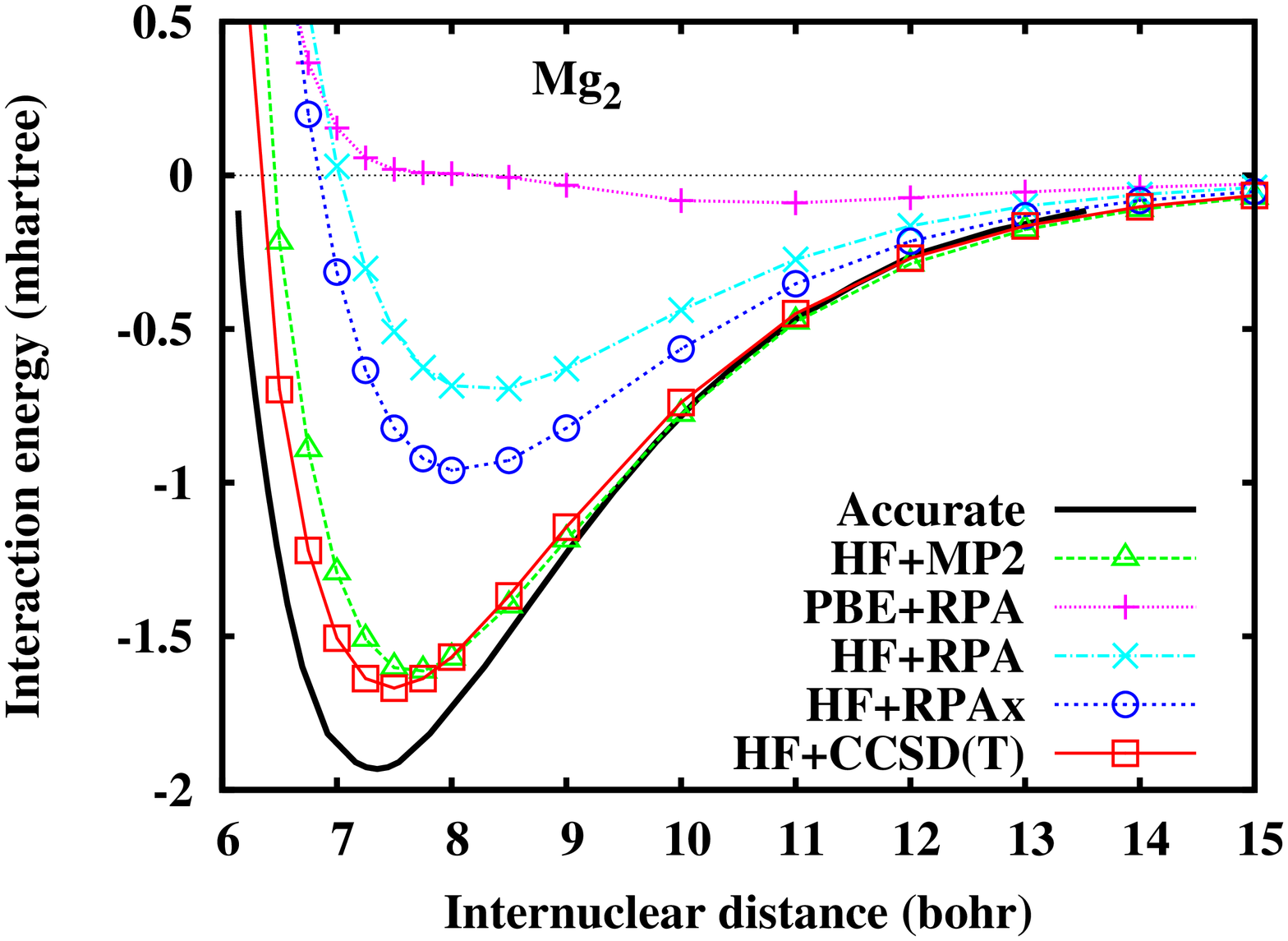}
\includegraphics[scale=0.30,angle=-90]{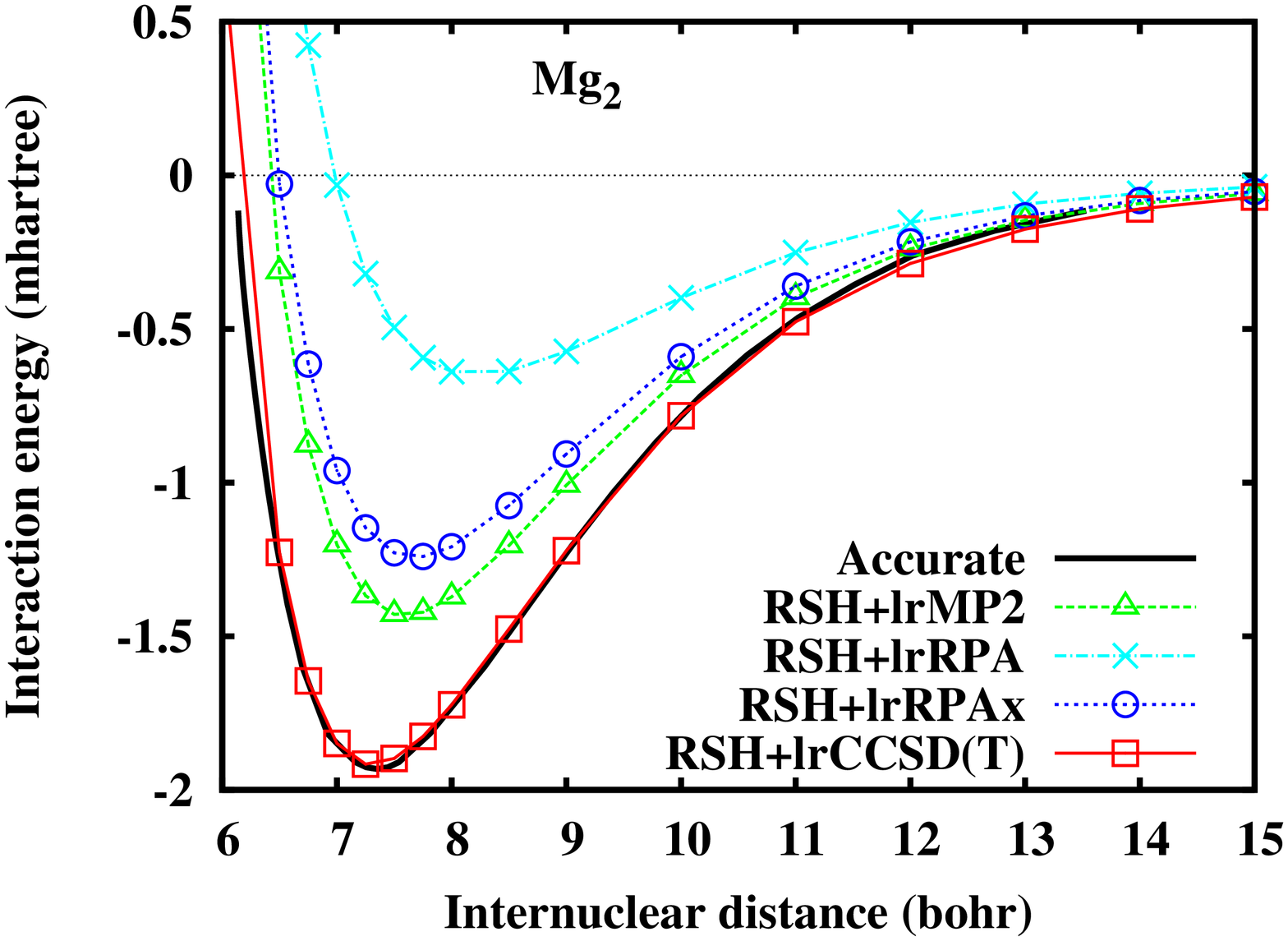}
\includegraphics[scale=0.30,angle=-90]{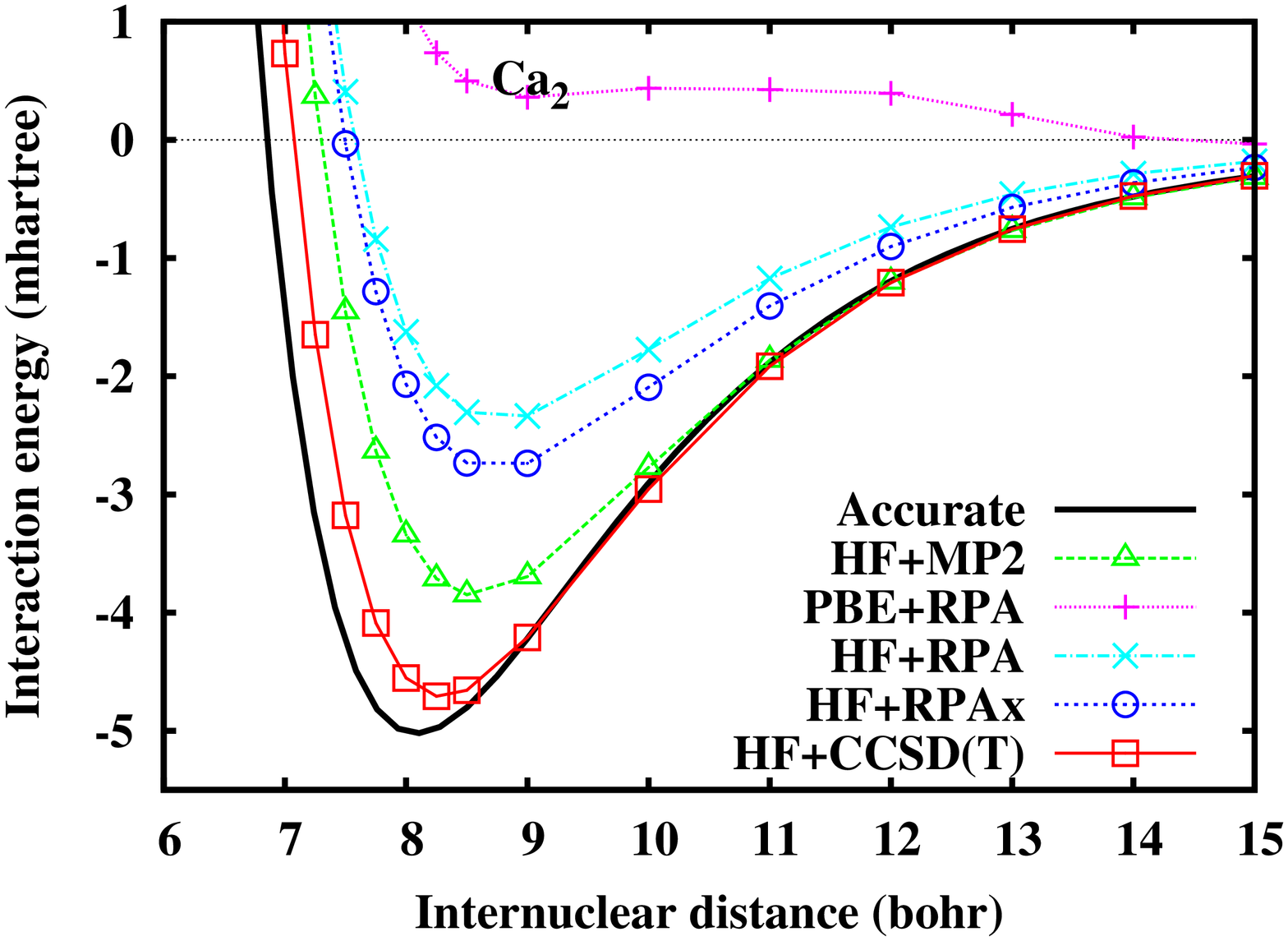}
\includegraphics[scale=0.30,angle=-90]{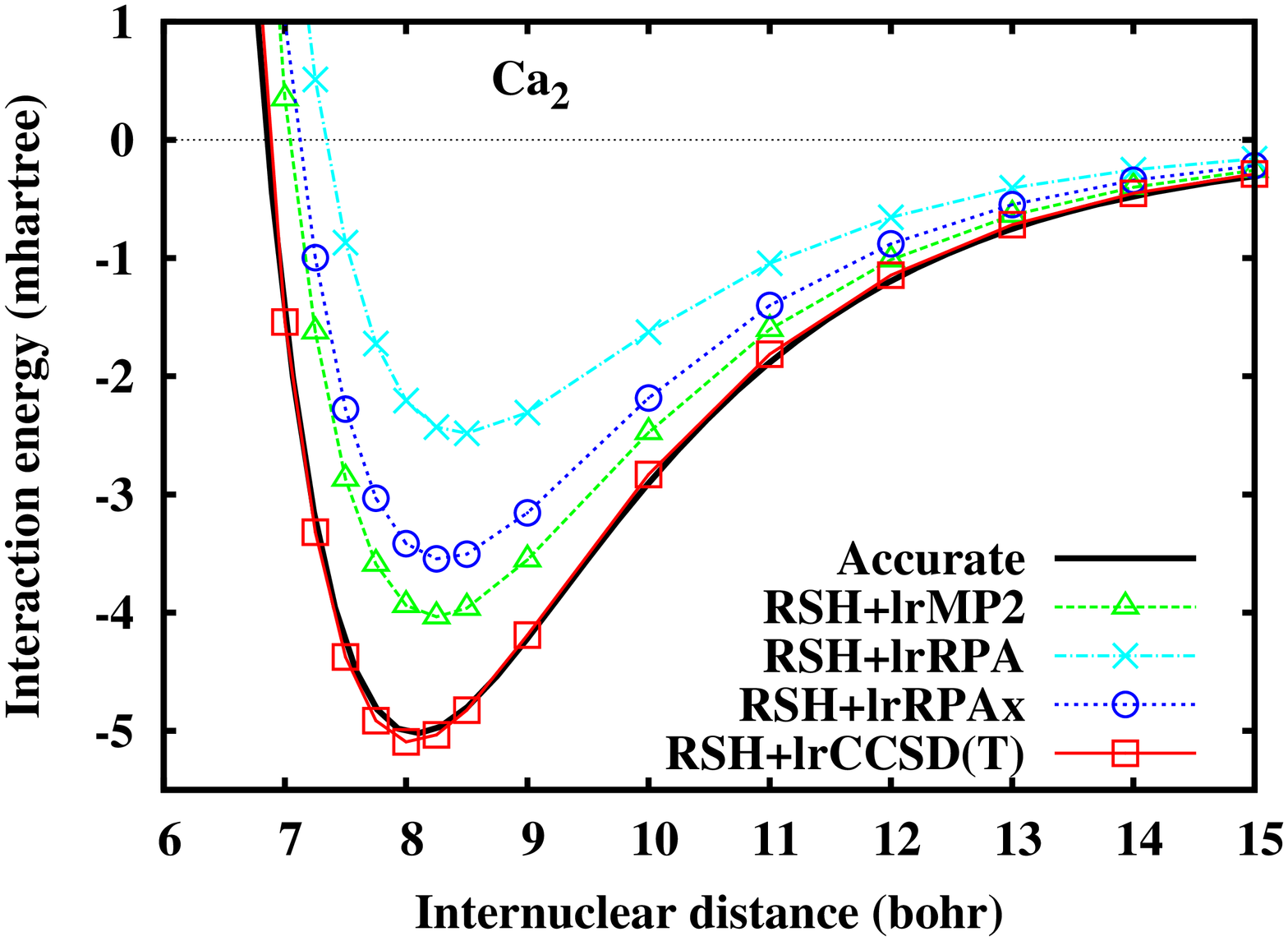}
\caption{(Color online) Interaction energy curves of Be$_2$, Mg$_2$ and Ca$_2$ calculated by full-range (left) and range-separated (right) methods. The basis is cc-pV5Z. The accurate curves are from Refs.~\onlinecite{RoeVes-IJQC-05},~\onlinecite{BalDou-CJP-70} and~\onlinecite{AllKnoTie-PRA-02}.
}
\label{fig:alkalineearth}
\end{figure*}

\begingroup
\squeezetable
\begin{table*}[t]
\caption{Hard-core radii $\sigma$ (bohr), equilibrium distances $R_e$ (bohr), equilibrium binding energies $D_e$ (mhartree), harmonic vibrational frequencies $\omega_e$ (cm$^{-1}$) and dispersion coefficients $C_6$ for Be$_2$, Mg$_2$ and Ca$_2$ from different full-range and range-separated methods with cc-pV5Z basis. Mean absolute percentage errors (MA\%E) are also given.}
\label{tab:alkalineearth}
\begin{tabular}{lrrrrrrrrrr} \hline \hline
          &  HF+MP2 & PBE+RPA    & HF+RPA &    HF+RPAx  & HF+CCSD(T)& RSH+lrMP2& RSH+lrRPA & RSH+lrRPAx & RSH+lrCCSD(T) & Estimated exact \\
\hline
\multicolumn{10}{l}{Be$_2$}\\
$\sigma$     &   4.44   &   4.34  &   5.59  &    5.30  &    4.16  &    4.25  &    4.50  &    4.27  &    3.87  &  4.01$^a$  \\   
$R_e$        &   5.15   &   4.60  &   7.48  &    7.17  &    4.71  &    4.92  &    5.08  &    4.92  &    4.54  &  4.63$^a$  \\   
$D_e$        &   1.92   &   0.58  &   0.39  &    0.56  &    2.70  &    2.95  &    1.24  &    2.81  &    6.92  &  4.31$^a$  \\   
$\omega_e$   &   139    &    297  &   34    &    37    &    242   &    199   &    152   &    198   &    315   &  267$^a$   \\   
$C_6$        &   256    &   164   &   138   &    180   &    195   &    232   &    149   &    213   &    274   &  214$^d$   \\   
\hline													                        
\multicolumn{10}{l}{Mg$_2$}\\										                        
$\sigma$     &   6.44   &   8.30  &   7.02  &    6.83  &    6.29  &    6.40  &    6.98  &    6.49  &    6.13  &  6.10$^b$  \\   
$R_e$        &   7.66   &  10.72  &   8.28  &    8.11  &    7.48  &    7.59  &    8.23  &    7.68  &    7.31  &  7.35$^b$  \\   
$D_e$        &   1.62   &   0.09  &   0.70  &    0.96  &    1.67  &    1.43  &    0.65  &    1.24  &    1.92  &  1.93$^b$  \\   
$\omega_e$   &    47    &    7.9  &   31    &      35  &      48  &      45  &      30  &      42  &      52  &  51.1$^b$  \\   
$C_6$        &   686    &    405  &   364   &     485  &     616  &     571  &     349  &     494  &     671  &   627$^d$  \\   
\hline													                        
\multicolumn{10}{l}{Ca$_2$}\\										                        
$\sigma$     &   7.29   &   ---   &   7.57  &    7.49  &    7.07  &    7.04  &    7.33  &    7.11  &    6.85  &  6.88$^c$  \\   
$R_e$        &   8.57   &   ---   &   8.76  &    8.72  &    8.30  &    8.25  &    8.47  &    8.30  &    8.05  &  8.09$^c$  \\   
$D_e$        &   3.85   &   ---   &   2.37  &    2.78  &    4.71  &    4.03  &    2.48  &    3.55  &    5.10  &  5.02$^c$  \\   
$\omega_e$   &   56     &   ---   &   44    &      47  &      64  &      60  &      50  &      57  &      68  &  63.7$^c$  \\   
$C_6$        &   2574   &  1335   &   1301  &    1710  &    2311  &    2090  &    1173  &    1617  &    2224  &   2221$^d$  \\  
\hline													                        
\multicolumn{10}{l}{MA\%E (\%)}\\									                        
$\sigma$     &   7.4    &   ---   &   22    &    18    &    3.2   &    4.4   &    11    &    5.4   &    1.5   &  0.0   \\       
$R_e$        &   7.1    &   ---   &   28    &    24    &    2.0   &    3.8   &    8.7   &    4.5   &    1.0   &  0.0   \\       
$D_e$        &   32     &   ---   &   69    &    61    &    19    &    26    &    63    &    33    &    21    &  0.0   \\       
$\omega_e$   &   23     &   ---   &   53    &    48    &    5.3   &    14    &    35    &    18    &    9.1   &  0.0   \\       
$C_6$        &   15     &    33   &   40    &    21    &    5.0   &    7.7   &    41    &    16    &    12    &  0.0   \\        
\hline\hline
$^a$ From Ref.~\onlinecite{RoeVes-IJQC-05}\\
$^b$ From Ref.~\onlinecite{BalDou-CJP-70}\\
$^c$ From Ref.~\onlinecite{AllKnoTie-PRA-02}\\
$^d$ From  Ref.~\onlinecite{PorDer-PRA-02}\\
\end{tabular}
\end{table*}
\endgroup

\subsection{Rare-gas dimers}

In Fig.~\ref{fig:raregas}, the interaction energy curves of He$_2$, Ne$_2$, Ar$_2$ and Kr$_2$, obtained with the full-range and range-separated methods are compared. As already known, full-range HF+MP2 underestimates the interaction energy for the smallest systems He$_2$ and Ne$_2$, and overestimates it for the largest systems Ar$_2$ and Kr$_2$. Full-range PBE+RPA gives an almost dissociative curve for He$_2$, and largely underestimates the interaction energy for Ne$_2$, Ar$_2$ and Kr$_2$. Using HF orbitals in full-range RPA drastically improves the interaction energy curve for He$_2$, and to a least extend for Ne$_2$, but gives less binding for Ar$_2$ and Kr$_2$. Full-range HF+RPAx significantly improves over full-range HF+RPA, but still gives underestimated interaction energies. It can be noted that full-range HF+RPAx yields interaction energy curves almost identical to the full-range HF+MP2 curves for He$_2$ and Ne$_2$, and almost identical to the full-range PBE+RPA curves for Ar$_2$ and Kr$_2$. Full-range HF+CCSD(T) gives systematically quite accurate interaction energies. Quite similarly to full-range HF+MP2, the range-separated RSH+lrMP2 underestimates the interaction energy for He$_2$ and Ne$_2$, and overestimates it for Ar$_2$ and Kr$_2$. RSH+lrRPA tends to improve over both full-range PBE+RPA and HF+RPA but still leads to significantly underestimated interaction energies. RSH+lrRPAx improves over both RSH+lrRPA and full-range HF+RPAx; it still systematically underestimates the interaction energy at equilibrium, but appears quite accurate at medium and large distances. On the contrary, RSH+lrCCSD(T) systematically overestimates the interaction energy at medium and large distances.

The hard-core radii, equilibrium distances, equilibrium binding energies, harmonic vibrational frequencies and dispersion coefficients $C_6$ for ten homonuclear and heteronuclear rare-gas dimers calculated with the full-range and range-separated methods are given in Table~\ref{tab:raregas}. The trends seen in Fig.~\ref{fig:raregas} are confirmed. Full-range RPA (with PBE or HF orbitals) yields very inaccurate equilibrium properties. Full-range HF+RPAx improves over full-range HF+RPA (with the exception of $C_6$ coefficients which turn out to be quite good in PBE+RPA for these systems) but the errors remain large. Range separation largely improves RPA and RPAx. RSH+lrRPAx gives much better equilibrium properties than RSH+lrRPA, with mean absolute percentage errors smaller by more than a factor of two, while these two methods give similar accuracy for $C_6$ coefficients. Full-range HF+MP2 is reasonably accurate and range separation has a much smaller impact on it. For these systems, RSH+lrMP2 gives an overall similar accuracy than RSH+RPAx, although the $C_6$ coefficients tend to be globally more accurate in RSH+lrRPAx. Full-range HF+CCSD(T) gives the best results. Surprisingly, range separation tends to deteriorate the accuracy of CCSD(T), especially for $C_6$ coefficients. Nevertheless, among the range-separated methods, RSH+lrCCSD(T) still gives the best equilibrium properties.


\subsection{Alkaline-earth dimers}

In Fig.~\ref{fig:alkalineearth}, the interaction energy curves of Be$_2$, Mg$_2$ and Ca$_2$, obtained with the full-range and range-separated methods are compared. These systems have static correlation effects, especially Be$_2$, and are thus more challenging for the single-reference methods tested here. Full-range PBE+RPA gives unphysical interaction energy curves, with a large bump for Be$_2$, and with essentially no bond for Mg$_2$ and Ca$_2$. Full-range HF+RPA yields an almost dissociative curve for Be$_2$ with no bump (which is consistent with Ref.~\onlinecite{NguGal-JCP-10}), and physically reasonable curves for Mg$_2$ and Ca$_2$. Full-range HF+RPAx moderately improves over full-range HF+RPA. Among the full-range methods, HF+MP2 and HF+CCSD(T) clearly give the best interaction energy curves. As for rare-gas dimers, RSH+lrRPA always largely underestimates the interaction energy. RSH+lrMP2 and RSH+lrRPAx give much less underestimated interaction energies, with RSH+lrMP2 being a bit more accurate for Mg$_2$ and Ca$_2$. While RSH+lrCCSD(T) largely overestimates the interaction energy for Be$_2$, it is remarkably accurate for Mg$_2$ and Ca$_2$. We note that RSH+lrCCSD(T) could be made more accurate for Be$_2$ by choosing a larger range-separation parameter $\mu$~\cite{ReiTouAngSav-JJJ-XX}.

The hard-core radii, equilibrium distances, equilibrium binding energies, harmonic vibrational frequencies and dispersion coefficients $C_6$ for Be$_2$, Mg$_2$ and Ca$_2$ are given in Table~\ref{tab:alkalineearth}. It is confirmed that range separation largely improves the equilibrium properties of RPA and RPAx. Again, RSH+lrRPAx is much more accurate than RSH+lrRPA, with mean absolute percentage errors smaller by about a factor of two. Range separation also overall brings a significant improvement in MP2. Among the range-separated methods, RSH+lrCCSD(T) gives the best equilibrium properties.



\section{Conclusions}

We have expounded the details of a formally exact adiabatic-connection fluctuation-dissipation density-functional theory based on range separation. Range-separated density-functional theory with random phase approximations including or not the long-range Hartree-Fock exchange response kernel (referred to as RSH+lrRPA and RSH+lrRPAx, respectively) are then obtained as well-identified approximations on the long-range Green-function self-energy [Eqs.~(\ref{ApproximationRPA}) and~(\ref{ApproximationRPAx})]. The long-range Green function does not vary along the adiabatic connection at the RSH+lrRPA and RSH+lrRPAx levels, which makes these schemes relatively simple compared to the exact theory. In practice, RSH+lrRPA and RSH+lrRPAx have been applied in a spin-restricted closed-shell formalism, in which both schemes only include spin-singlet orbital excitations, and thus are not subject to triplet instabilities. 

These range-separated RPA-type schemes have been tested on rare-gas and alkaline-earth dimers, featuring challenging weak (van der Waals) interactions. Both range separation and inclusion of the exact Hartree-Fock response kernel largely improve the accuracy of RPA. The RSH+lrRPAx method appears as a reasonably accurate method for weak interactions, but globally less accurate for equilibrium properties than the more intensive range-separated coupled-cluster method. Although, for the small systems considered here, range-separated second-order perturbation theory (RSH+lrMP2) turns out to yield results similarly as accurate as those from RSH+lrRPAx (and in fact more accurate for Mg$_2$ and Ca$_2$), a recent investigation~\cite{ZhuTouSavAng-JCP-10} shows that RSH+lrRPAx corrects the overestimation of the binding energy in RSH+lrMP2 for larger weakly-interacting stacked complexes, such as the benzene dimer.

\section*{Acknowledgments}

We thank J. F. Dobson and T. Gould for numerous discussions on RPA during the FAST (French-Australian Science and Technology) workshop at Griffith University, Australia. We also thank G. Jansen for discussions. This work was partly supported by ANR (French national research agency) via contract number 07-BLAN-0272 (Wademecom).

\appendix
\section{Adiabatic-connection fluctuation-dissipation density-functional theory}
\label{app:manybody}

In this appendix, we outline a general, formally exact adiabatic-connection fluctuation-dissipation density-functional theory, using Green-function many-body theory. For further details on standard Green's function theory, see e.g. Refs.~\onlinecite{Str-RNC-88,GroRunHei-BOOK-91,OniReiRub-RMP-02,Bru-THESIS-05}.

\subsection{Adiabatic connection}

We consider the following adiabatic connection defined by the $\lambda$-dependent energy
\begin{equation}
E_\l = \min_\Psi \left\{ \bra{\Psi} \hat{K}_0 + \l \hat{W} \ket{\Psi} + F[n_\Psi] \right\},
\end{equation}
where $\hat{K}_0$ is an arbitrary one-particle Hamiltonian, $\hat{W}$ is a perturbation operator (generally, the sum of a two-particle operator $\hat{W}_{ee}$ and an one-particle operator) and $F[n]$ is a $\lambda$-independent density functional. The minimizing multideterminant wave function $\Psi_\l$ satisfies the Euler-Lagrange equation
\begin{equation}
\hat{H}_\l \ket{\Psi_\l} = {\cal E}_\l \ket{\Psi_\l},
\end{equation}
where ${\cal E}_\l$ is the Lagrange multiplier for the normalization constraint, and $\hat{H}_\l$ is the effective Hamiltonian along the adiabatic connection
\begin{equation}
\hat{H}_\l = \hat{K}_0 + \l \hat{W} + \hat{V}_\l,
\end{equation}
where $\hat{V}_\l = \int d\b{r} \, \delta F[n_{\Psi_\l}]/\delta n(\b{r}) \, \hat{n}(\b{r})$ is a self-consistent one-particle potential operator. Note that $\hat{H}_{\l=1}$ is not necessarily the physical Hamiltonian. This adiabatic connection links the energy of interest $E_{\l=1}$ to the reference energy $E_{\l=0}=\bra{\Phi_0} \hat{K}_0 \ket{\Phi_0} + F[n_{\Phi_0}]$ calculated with the single-determinant wave function $\Phi_0=\Psi_{\l=0}$ of the reference Hamiltonian $\hat{H}_0 = \hat{K}_0 + \hat{V}_0$. The one-particle density is not kept constant with respect to $\l$.

An adiabatic connection formula for $E_{\l=1}$ is found by taking the derivative of $E_{\l}$ with respect to $\l$, noting that $E_{\l}$ is stationary with respect to $\Psi_{\l}$, and reintegrating between $\l=0$ and $\l=1$
\begin{eqnarray}
E_{\l=1} = E_{\l=0} + \int_{0}^{1} d\l \, \bra{\Psi_\l} \hat{W}  \ket{\Psi_\l}.
\label{}
\end{eqnarray}
The correlation energy, defined as $E_c = E_{\l=1} - E_{\l=0} - (d E_\l/d\l)_{\l=0}$ where $(d E_\l/d\l)_{\l=0}=\bra{\Phi_0} \hat{W}  \ket{\Phi_0}$ is the first-order energy correction, is thus given by
\begin{eqnarray}
E_c = \int_{0}^{1} d\l \left[ \bra{\Psi_\l} \hat{W}  \ket{\Psi_\l} - \bra{\Phi_0} \hat{W} \ket{\Phi_0} \right],
\label{}
\end{eqnarray}
or, equivalently, in the representation of space-spin coordinates $\b{x}=(\b{r},s)$
\begin{eqnarray}
E_c = \frac{1}{2} \int_{0}^{1} d\l \int d\b{x}_1 d\b{x}_2 d\b{x}_1' d\b{x}_2' w(\b{x}_1,\b{x}_2;\b{x}_1',\b{x}_2') 
\nonumber\\
\times P_{c,\l}(\b{x}_1,\b{x}_2;\b{x}_1',\b{x}_2'),
\label{Ecspacespin}
\end{eqnarray}
where $w(\b{x}_1,\b{x}_2;\b{x}_1',\b{x}_2')$ is the interaction potential corresponding to the operator $\hat{W}$ and $P_{c,\l}(\b{x}_1,\b{x}_2;\b{x}_1',\b{x}_2')$ is the correlation part of the two-particle density matrix along the adiabatic connection.

This exposition encompasses both standard full-range many-body theory and range-separated density-functional theory. Indeed, if $\hat{K}_0$ is the Hartree-Fock Hamiltonian (i.e., $\hat{K}_0=\hat{T}+\hat{V}_{ne}+\hat{V}_{\H x,\HF}$), $\hat{W}$ is the standard M{\o}ller-Plesset fluctuation perturbation operator (i.e., $\hat{W}=\hat{W}_{ee}-\hat{V}_{\H x,\HF}$) and $F[n]=0$ then Eq.~(\ref{Ecspacespin}) yields the full-range many-body correlation energy, defined with respect to the Hartree-Fock energy. Similarly, with the corresponding long-range operators $\hat{K}_0=\hat{T}+\hat{V}_{ne}+\hat{V}^\lr_{\H x,\HF}$ and $\hat{W}=\hat{W}_{ee}^\lr-\hat{V}_{\H x,\HF}^\lr$ and the short-range density functional $F[n]=E_{\H xc}^\sr[n]$, Eq.~(\ref{Ecspacespin}) yields now the long-range correlation energy, defined with respect to the RSH energy [Eq.~(\ref{ERSH+Eclr})].

\subsection{One-particle Green function}

The one-particle Green function along the adiabatic connection is defined as
\begin{equation}
G_\l(1,2)= -i \bra{\Psi_\l} T [ \hat{\psi}_\l(1) \hat{\psi}_\l^\dag(2) ] \ket{\Psi_\l},
\end{equation}
where $1=(\b{x}_1,t_1)$ and $2=(\b{x}_2,t_2)$ refer to space-spin and time coordinates, $\hat{\psi}_\l(1) = e^{i \hat{H}_\l t_1} \hat{\psi}(\b{x}_1) e^{-i \hat{H}_\l t_1}$ and $\hat{\psi}_\l^\dag(2) = e^{i \hat{H}_\l t_2} \hat{\psi}^\dag(\b{x}_2) e^{-i \hat{H}_\l t_2}$ are the annihilation and creation operators in the Heisenberg picture, and $T$ is the Wick time-ordering operator.

A Dyson-type equation connects the inverse of $G_\l$ to the inverse of the Green function associated with the one-electron Hamiltonian $\hat{K}_0+\hat{V}_\l$, denoted by $G_{V,\l}$, 
\begin{equation}
G_\l^{-1}(1,2) = G_{V,\l}^{-1}(1,2) - \Sigma_\l(1,2),
\label{Dysoneq}
\end{equation}
which can be considered as the definition of the self-energy $\Sigma_\l$. In turn, the inverse of $G_{V,\l}$ can be expressed from the inverse of the Green function $G_{0}$ of the reference Hamiltonian $\hat{H}_0=\hat{K}_0+\hat{V}_0$ as $G_{V,\l}^{-1} = G_{0}^{-1} - [v_\l-v_0]$, where $v_\l$ and $v_0$ are the one-electron potentials associated with $\hat{V}_\l$ and $\hat{V}_0$, respectively.

For time-independent Hamiltonians, the Green function only depends on the time difference $\tau=t_1-t_2$, so one defines $G_\l(\b{x}_1,\b{x}_2;\tau)=G_\l(\b{x}_1 t_1,\b{x}_2 t_2)$, which has a discontinuity at $\tau=0$. The one-particle density matrix $n_{1,\l}(\b{x}_1,\b{x}_2)=\bra{\Psi_\l} \hat{n}_1(\b{x}_1,\b{x}_2) \ket{\Psi_\l}$, with $\hat{n}_1(\b{x}_1,\b{x}_2)=\hat{\psi}^\dag(\b{x}_2) \hat{\psi}(\b{x}_1)$, can be obtained from the limit $\tau \to 0^-$
\begin{equation}
n_{1,\l}(\b{x}_1,\b{x}_2) = -i G_\l(\b{x}_1,\b{x}_2;\tau=0^-).
\label{n1lG}
\end{equation}

\subsection{Four-point polarization propagator}
\label{sec:4ptpolpropa}

The four-point polarization propagator along the adiabatic connection is defined as
\begin{equation}
\chi_\l(1,2;1',2')= i \left[ G_{2,\l}(1,2;1',2') - G_\l(1,1') G_\l(2,2')\right],
\end{equation}
where $G_{2,\l}$ is the two-particle Green function
\begin{equation}
G_{2,\l}(1,2;1',2')= - \bra{\Psi_\l} T [ \hat{\psi}_\l(1) \hat{\psi}_\l(2) \hat{\psi}_\l^\dag(2') \hat{\psi}_\l^\dag(1') ] \ket{\Psi_\l},
\label{G2l}
\end{equation}
Alternatively, using the Schwinger derivative technique, $\chi_\l$ can be expressed as the functional derivative of the one-particle Green function with respect to the two-point potential $v_\l$ (see, e.g., Refs.~\onlinecite{Str-RNC-88,Bru-THESIS-05})
\begin{equation}
\chi_\l(1,2;1',2')= -i \frac{\delta G_{V,\l}(1,1')}{\delta v_\l(2',2)}.
\label{childefderiv}
\end{equation}
The four-point polarization propagator satisfies a so-called Bethe-Salpeter equation that directly stems from the Dyson equation of Eq.~(\ref{Dysoneq}). Considering variations with respect to $i G_\l$ (achieved through variations of $v_\l$) yields
\begin{equation}
-i \frac{\delta G_\l^{-1}(1,1')}{\delta G_\l(2',2)} = -i \frac{\delta G_{V,\l}^{-1}(1,1')}{\delta G_\l(2',2)} + i \frac{\delta \Sigma_\l(1,1')}{\delta G_\l(2',2)}.
\label{DysoneqUderiv}
\end{equation}
The term on the left-hand side of Eq.~(\ref{DysoneqUderiv}) gives straightforwardly
\begin{eqnarray}
-i\frac{\delta G_\l^{-1}(1,1')}{\delta G_\l(2',2)} &=& i G_\l^{-1}(1,2') G_\l^{-1}(2,1')
\nonumber\\
&=& \chi_{\IP,\l}^{-1}(1,2;1',2'),
\label{ChiIPl-1}
\end{eqnarray}
where $\chi_{\IP,\l}(1,2;1',2')=-i G_\l(1,2') G_\l(2,1')$ is a so-called independent-particle (IP) polarization propagator~\cite{TouZhuAngSav-JJJ-XX-note3}. The first term on the right-hand side of Eq.~(\ref{DysoneqUderiv}) gives the inverse of the four-point polarization propagator, according to Eq.~(\ref{childefderiv}),
\begin{eqnarray}
-i \frac{\delta G_{V,\l}^{-1}(1,1')}{\delta G_\l(2',2)} &=& i \frac{\delta v_\l(1,1')}{\delta G_\l(2',2)} 
= \chi_\l^{-1}(1,2;1',2'), \,\,\,\,\,
\end{eqnarray}
and the second term is the so-called Bethe-Salpeter four-point kernel
\begin{eqnarray}
i\frac{\delta \Sigma_\l(1,1')}{\delta G_\l(2',2)} = f_\l(1,2;1',2'),
\label{fl}
\end{eqnarray}
and finally, using Eqs.~(\ref{ChiIPl-1})-(\ref{fl}) in Eq.~(\ref{DysoneqUderiv}), the Bethe-Salpeter equation for $\chi_\l$ writes
\begin{equation}
\chi_\l^{-1}(1,2;1',2') = \chi_{\IP,\l}^{-1}(1,2;1',2') - f_\l(1,2;1',2').
\label{Bethe-Salpeter}
\end{equation}

\subsection{Fluctuation-dissipation theorem}

Similarly to the expression of the one-particle density matrix in terms of the one-particle Green function [Eq.~(\ref{n1lG})], the two-particle density matrix can be extracted from the polarization propagator. Defining $\chi_\l(\b{x}_1,\b{x}_2;\b{x}_1',\b{x}_2';\tau) = \chi_\l(\b{x}_1 t_1,\b{x}_2 t_2;\b{x}_1' t_1^+,\b{x}_2't_2^+)$, i.e. the polarization propagator with times $t_1' \to t_1^+$ and $t_2' \to t_2^+$ which depends only on the time difference $\tau=t_1-t_2$, it is easy to check that in the limit $\tau \to 0^-$, after applying the time-ordering operator in Eq.~(\ref{G2l}) and using Eq.~(\ref{n1lG}), one has the following relation
\begin{eqnarray}
i\chi_\l(\b{x}_1,\b{x}_2;\b{x}_1',\b{x}_2';\tau=0^-) &=& \bra{\Psi_\l} \hat{n}_1(\b{x}_2,\b{x}_2') \hat{n}_1(\b{x}_1,\b{x}_1')\ket{\Psi_\l} 
\nonumber\\
&&-n_{1,\l}(\b{x}_1,\b{x}_1')n_{1,\l}(\b{x}_2,\b{x}_2').
\nonumber\\
\end{eqnarray}
The two-particle density matrix $n_{2,\l}(\b{x}_1,\b{x}_2;\b{x}_1',\b{x}_2') = \bra{\Psi_\l} \hat{\psi}^\dag(\b{x}_2') \hat{\psi}^\dag(\b{x}_1') \hat{\psi}(\b{x}_1) \hat{\psi}(\b{x}_2) \ket{\Psi_\l}$ can thus be expressed as
\begin{eqnarray}
n_{2,\l}(\b{x}_1,\b{x}_2;\b{x}_1',\b{x}_2') &=&  \bra{\Psi_\l} \hat{n}_1(\b{x}_2,\b{x}_2') \hat{n}_1(\b{x}_1,\b{x}_1')\ket{\Psi_\l} 
\nonumber\\
&&-\delta(\b{x}_1'-\b{x}_2) n_{1,\l}(\b{x}_1,\b{x}_2')
\nonumber\\
&=& i\chi_\l(\b{x}_1,\b{x}_2;\b{x}_1',\b{x}_2';\tau=0^-)
\nonumber\\
&&+n_{1,\l}(\b{x}_1,\b{x}_1')n_{1,\l}(\b{x}_2,\b{x}_2')
\nonumber\\
&&-\delta(\b{x}_1'-\b{x}_2) n_{1,\l}(\b{x}_1,\b{x}_2').
\end{eqnarray}
The correlation part of the two-particle density matrix $P_{c,\l}=n_{2,\l}-n_{2,\l=0}$ is thus
\begin{eqnarray}
P_{c,\l}(\b{x}_1,\b{x}_2;\b{x}_1',\b{x}_2') &=& i\chi_\l(\b{x}_1,\b{x}_2;\b{x}_1',\b{x}_2';\tau=0^-)
\nonumber\\
&&-i\chi_0(\b{x}_1,\b{x}_2;\b{x}_1',\b{x}_2';\tau=0^-)
\nonumber\\
&&+\Delta_\l(\b{x}_1,\b{x}_2;\b{x}_1',\b{x}_2'),
\end{eqnarray}
where $\chi_0$ is the polarization propagator of the non-interacting reference system for $\l=0$, and $\Delta_\l$ is a term coming from the variation of the one-particle density matrix along the adiabatic connection
\begin{eqnarray}
\Delta_\l(\b{x}_1,\b{x}_2;\b{x}_1',\b{x}_2') &=& n_{1,\l}(\b{x}_1,\b{x}_1')n_{1,\l}(\b{x}_2,\b{x}_2')
\nonumber\\
&&-\delta(\b{x}_1'-\b{x}_2) n_{1,\l}(\b{x}_1,\b{x}_2')
\nonumber\\
&&-n_{1,0}(\b{x}_1,\b{x}_1')n_{1,0}(\b{x}_2,\b{x}_2')
\nonumber\\
&&+\delta(\b{x}_1'-\b{x}_2) n_{1,0}(\b{x}_1,\b{x}_2').
\label{deltal}
\end{eqnarray}
Using Eq.~(\ref{n1lG}), one can also express this term with the Green function as $\Delta_\l=\Gamma[G_{\l}]-\Gamma[G_{0}]$ where we define the functional $\Gamma$ as
\begin{eqnarray}
\Gamma[G]&=& - G(\b{x}_1,\b{x}_1';\tau=0^-) G(\b{x}_2,\b{x}_2';\tau=0^-)
\nonumber\\
&&+\delta(\b{x}_1'-\b{x}_2) i G(\b{x}_1,\b{x}_2';\tau=0^-).
\label{GammaG}
\end{eqnarray}
Finally, introducing the Fourier transform of $\chi_\l(\b{x}_1,\b{x}_2;\b{x}_1',\b{x}_2';\tau)$ in terms of the frequency $\omega$,
\begin{eqnarray}
i\chi_\l(\b{x}_1,\b{x}_2;\b{x}_1',\b{x}_2';\tau=0^-) = - \int_{-\infty}^{\infty} \frac{d\omega}{2\pi i} e^{i\omega 0^+} 
\nonumber\\
\times \chi_\l(\b{x}_1,\b{x}_2;\b{x}_1',\b{x}_2';\omega),
\end{eqnarray}
we arrive at the form of the fluctuation-dissipation that we use
\begin{eqnarray}
P_{c,\l}(\b{x}_1,\b{x}_2;\b{x}_1',\b{x}_2') = - \int_{-\infty}^{\infty} \frac{d\omega}{2\pi i} e^{i\omega 0^+} \Bigl[ \chi_\l(\b{x}_1,\b{x}_2;\b{x}_1',\b{x}_2';\omega)
\nonumber\\
-\chi_0(\b{x}_1,\b{x}_2;\b{x}_1',\b{x}_2';\omega) \Bigl] +\Delta_\l(\b{x}_1,\b{x}_2;\b{x}_1',\b{x}_2'). \;\;\;\;\;\;\;
\label{Pcl}
\end{eqnarray}

\section{Random phase approximation in an orbital basis}
\label{app:rpabasis}

In this appendix, we give the working equations in an orbital basis resulting from the many-body theory outlined in Appendix~\ref{app:manybody}, in the special case of a random phase approximation (RPA)-type simplification. For further details, see e.g. Refs.~\onlinecite{McW-BOOK-92,MclBal-RMP-64,Fur-JCP-01,Fur-PRB-01}.

\subsection{Expressions in a spin-orbital basis}

In the RPA and RPAx approximations, the Green function does not vary along the adiabatic connection, i.e. $G_\l=G_0$, which implies that the independent-particle polarization propagator [Eq.~(\ref{ChiIPl-1})] is just the non-interacting reference polarization propagator, $\chi_{\IP,\l}(1,2;1',2')=-i G_0(1,2') G_0(2,1')=\chi_0(1,2;1',2')$, and in the fluctuation-dissipation theorem of Eq.~(\ref{Pcl}) the term coming from the variation of the one-particle density matrix vanishes, $\Delta_\l=0$.

The frequency-dependent non-interacting polarization propagator has the following well-known Lehmann representation
\begin{eqnarray}
\chi_0(\b{x}_1,\b{x}_2;\b{x}_1',\b{x}_2';\omega) = \sum_{ia} \frac{\phi_i^*(\b{x}_1') \phi_a(\b{x}_1) \phi_a^*(\b{x}_2') \phi_i(\b{x}_2)}{\omega-(\epsilon_a-\epsilon_i)+i0^+}
\nonumber\\
- \sum_{ia} \frac{\phi_i^*(\b{x}_2') \phi_a(\b{x}_2) \phi_a^*(\b{x}_1') \phi_i(\b{x}_1)}{\omega+(\epsilon_a-\epsilon_i)-i0^+}, \,\,\,\,\,\,
\label{}
\end{eqnarray}
where $\phi_p(\b{x})$ and $\epsilon_p$ are the spin orbitals and corresponding eigenvalues of the reference system, and $i$ and $a$ run over occupied and virtual spin orbitals, respectively. Hence, $\chi_0$ can be completely represented in the basis of spin-orbital products, $\phi_p^*(\b{x}_1') \phi_q(\b{x}_1)$, where $p$ refer to an occupied orbital and $q$ to a virtual orbital, and vice versa, with matrix elements
\begin{eqnarray}
\left( \mathbb{\Pi}_0 (\omega) \right)_{pq,rs} = \int d\b{x}_1 d\b{x}_2 d\b{x}_1' d\b{x}_2' \phi_p(\b{x}_1') \phi_q^*(\b{x}_1)
\nonumber\\
\times \chi_0(\b{x}_1,\b{x}_2;\b{x}_1',\b{x}_2';\omega) \phi_r^*(\b{x}_2) \phi_s(\b{x}_2').
\label{}
\end{eqnarray}
Assuming orthonormality of the spin orbitals, the matrix elements are easily calculated
\begin{subequations}
\begin{eqnarray}
\left( \mathbb{\Pi}_0 (\omega) \right)_{ia,jb} &=& \frac{\delta_{ij} \delta_{ab}}{\omega-(\epsilon_a-\epsilon_i)+i0^+},
\end{eqnarray}
\begin{eqnarray}
\left( \mathbb{\Pi}_0 (\omega) \right)_{ai,bj} &=& -\frac{\delta_{ij} \delta_{ab}}{\omega+(\epsilon_a-\epsilon_i)-i0^+},
\end{eqnarray}
\begin{eqnarray}
\left( \mathbb{\Pi}_0 (\omega) \right)_{ai,jb} &=& \left( \mathbb{\Pi}_0 (\omega) \right)_{ia,bj} = 0,
\end{eqnarray}
\label{Pi0elements}
\end{subequations}
where both $i$ and $j$ refer to occupied orbitals and both $a$ and $b$ to virtual orbitals. The matrix is thus diagonal, and the inverse of $\chi_0$ has the following $2\times2$ supermatrix representation
\begin{eqnarray}
\mathbb{\Pi}_0 (\omega)^{-1} &=& - \left[ \left( \begin{array}{cc} \bm{\Delta\epsilon} & \b{0} \\  \b{0} & \bm{\Delta\epsilon}\\ \end{array} \right) - 
\omega  \left( \begin{array}{cc} \b{1} & \b{0}\\  \b{0} & -\b{1}\\ \end{array} \right) \right],
\label{}
\end{eqnarray}
where $\bm{\Delta\epsilon}_{ia,jb} = (\epsilon_a-\epsilon_i) \delta_{ij} \delta_{ab}$, each block matrices being re-indexed with the composite indices $ia$ and $jb$.

In the RPA and RPAx approximations, the Bethe-Salpeter kernel of Eq.~(\ref{fl}) is approximated as the frequency-independent Hartree(-Fock) form [Eqs.~(\ref{fHlr}) and~(\ref{fxlr})]
\begin{eqnarray}
f_\l (\b{x}_1,\b{x}_2;\b{x}_1',\b{x}_2') &=& \l w_{ee}(r_{12}) [ \delta(\b{x}_1-\b{x}_1') \delta(\b{x}_2-\b{x}_2')
\nonumber\\
&&- \xi \, \delta(\b{x}_1-\b{x}_2')\delta(\b{x}_1'-\b{x}_2) ],
\label{}
\end{eqnarray}
where $w_{ee}(r_{12})$ is a two-particle interaction, and $\xi=0$ or $\xi=1$ for RPA and RPAx, respectively. This kernel has the following supermatrix elements
\begin{eqnarray}
\left( \mathbb{F}_\l \right)_{pq,rs} &=& \int d\b{x}_1 d\b{x}_2 d\b{x}_1' d\b{x}_2' \phi_p(\b{x}_1') \phi_q^*(\b{x}_1) 
\nonumber\\
&&\times f_\l(\b{x}_1,\b{x}_2;\b{x}_1',\b{x}_2') \phi_r^*(\b{x}_2) \phi_s(\b{x}_2')
\nonumber\\
&=& \l \left[ \bra{q r} \hat{w}_{ee} \ket{p s} - \xi \bra{q r} \hat{w}_{ee} \ket{s p} \right],
\label{}
\end{eqnarray}
where $\bra{q r} \hat{w}_{ee} \ket{p s}$ are the two-electron integrals. The supermatrix representation of the interacting polarization propagator $\chi_\l$ is then found from the Bethe-Salpeter equation [Eq.~(\ref{Bethe-Salpeter})] written in the spin-orbital basis
\begin{eqnarray}
\mathbb{\Pi}_\l (\omega)^{-1} &=& \mathbb{\Pi}_0 (\omega)^{-1} - \mathbb{F}_\l 
\nonumber\\
&=& - \left[ \left( \begin{array}{cc} \b{A}_\l & \b{B}_\l \\  \b{B}_\l^* & \b{A}_\l^* \\ \end{array} \right) - 
\omega  \left( \begin{array}{cc} \b{1} & \b{0}\\  \b{0} & -\b{1}\\ \end{array} \right) \right],
\label{}
\end{eqnarray}
where $\b{A}_\l$ and $\b{B}_\l$ are the so-called orbital rotation Hessians
\begin{subequations}
\begin{eqnarray}
\left( \b{A}_\l \right)_{ia,jb} &=&  (\epsilon_a-\epsilon_i) \delta_{ij} \delta_{ab} 
\nonumber\\
&&+ \l \left [ \bra{i b} \hat{w}_{ee} \ket{a j} - \xi \bra{i b} \hat{w}_{ee} \ket{j a} \right],
\label{}
\end{eqnarray}
\begin{eqnarray}
\left( \b{B}_\l \right)_{ia,jb} =  \l \left [ \bra{a b} \hat{w}_{ee} \ket{i j} - \xi \bra{a b} \hat{w}_{ee} \ket{j i} \right].
\label{}
\end{eqnarray}
\end{subequations}
We need to consider the linear response non-Hermitian eigenvalue equation
\begin{eqnarray}
\left( \begin{array}{cc} \b{A}_\l & \b{B}_\l \\  \b{B}_\l^* & \b{A}_\l^* \\ \end{array} \right) \left( \begin{array}{c} \b{X}_{n,\l} \\ \b{Y}_{n,\l} \\ \end{array} \right) =  \omega_{n,\l}  \left( \begin{array}{cc} \b{1} & \b{0}\\  \b{0} & -\b{1}\\ \end{array} \right) \left( \begin{array}{c} \b{X}_{n,\l} \\ \b{Y}_{n,\l} \\ \end{array} \right),
\label{eigeneq}
\end{eqnarray}
whose solutions come in pairs: positive excitation energies $\omega_{n,\l}$ with eigenvectors $\left( \b{X}_{n,\l}, \b{Y}_{n,\l} \right)$, and opposite (de-)excitation energies $-\omega_{n,\l}$ with eigenvectors $\left( \b{Y}_{n,\l}^*, \b{X}_{n,\l}^* \right)$. Choosing the normalization of the eigenvectors so that $\b{X}_{n,\l}^\dag \b{X}_{m,\l} - \b{Y}_{n,\l}^\dag \b{Y}_{m,\l}=\delta_{n m}$, the supermatrix $\mathbb{\Pi}_\l (\omega)$ can be expressed as the following spectral representation (where the sum is over eigenvectors with positive excitation energies)
\begin{eqnarray}
\mathbb{\Pi}_\l (\omega) = \sum_n \Biggl[ \frac{1}{\omega -\omega_{n,\l} +i0^+} \left( \begin{array}{c} \b{X}_{n,\l} \\ \b{Y}_{n,\l} \\ \end{array} \right) \left( \begin{array}{cc} \b{X}_{n,\l}^\dag & \b{Y}_{n,\l}^\dag \\ \end{array} \right) \\
\nonumber\\
- \frac{1}{\omega +\omega_{n,\l} -i0^+} \left( \begin{array}{c} \b{Y}_{n,\l}^* \\ \b{X}_{n,\l}^* \\ \end{array} \right) \left( \begin{array}{cc} \b{Y}_{n,\l}^{*\dag} & \b{X}_{n,\l}^{*\dag} \\ \end{array} \right) \Biggl].
\label{}
\end{eqnarray}

The fluctuation-dissipation theorem [Eq.~(\ref{Pcl})] leads to the supermatrix representation of the correlation part of the two-particle density matrix $P_{c,\l}$ (using contour integration in the upper half of the complex plane)
\begin{eqnarray} 
\mathbb{P}_{c,\l} &=& - \int_{-\infty}^{\infty} \frac{d\omega}{2\pi i} e^{i\omega0^+} [ \mathbb{\Pi}_\l (\omega) - \mathbb{\Pi}_0 (\omega)]
\nonumber\\
&=& \sum_n \left( \begin{array}{cc} \b{Y}_{n,\l}^* \b{Y}_{n,\l}^{*\dag} & \b{Y}_{n,\l}^* \b{X}_{n,\l}^{*\dag} \\ \b{X}_{n,\l}^* \b{Y}_{n,\l}^{*\dag} & \b{X}_{n,\l}^* \b{X}_{n,\l}^{*\dag}\\ \end{array} \right) - \left( \begin{array}{cc} \b{0} & \b{0}\\  \b{0} & \b{1}\\ \end{array} \right),
\label{}
\end{eqnarray}
the simple contribution coming from $\mathbb{\Pi}_0 (\omega)$ resulting from its diagonal form [Eqs.~(\ref{Pi0elements})], and the correlation energy [Eq.~(\ref{Ecspacespin})] has the following expression in spin-orbital basis
\begin{eqnarray} 
E_c &=& \frac{1}{2} \int_{0}^{1} d\l \sum_{pq,rs} \bra{p s} \hat{w} \ket{q r} (\mathbb{P}_{c,\l})_{pq,rs}
\nonumber\\
    &=& \frac{1}{2} \int_{0}^{1} d\l \sum_{ia,jb} \sum_n \Biggl\{ 
  \bra{i b} \hat{w}_{ee} \ket{a j} (\b{Y}_{n,\l})_{ia}^* (\b{Y}_{n,\l})_{jb}
\nonumber\\
&&+ \bra{i j} \hat{w}_{ee} \ket{a b} (\b{Y}_{n,\l})_{ia}^* (\b{X}_{n,\l})_{jb} + \bra{a b} \hat{w}_{ee} \ket{i j} (\b{X}_{n,\l})_{ia}^* (\b{Y}_{n,\l})_{jb}
\nonumber\\
&&+ \bra{a j} \hat{w}_{ee} \ket{i b} \left[(\b{X}_{n,\l})_{ia}^* (\b{X}_{n,\l})_{jb} -\delta_{ij}\delta_{ab} \right] \Biggl\},
\end{eqnarray}
where out of the integrals $\bra{p s} \hat{w} \ket{q r}$ associated with the general perturbation operator only the integrals of the type $\bra{i b} \hat{w}_{ee} \ket{a j}$ associated with the two-electron contribution of the perturbation operator survive because of the occupied-virtual/occupied-virtual structure of the two-particle density matrix. Using now real spin orbitals, the correlation energy can be simplified to
\begin{eqnarray} 
E_c &=& \frac{1}{2} \int_{0}^{1} d\l \sum_{ia,jb} \bra{i b} \hat{w}_{ee} \ket{a j} (\b{P}_{c,\l})_{ia,jb},
\label{Ecspinorbitals}
\end{eqnarray}
where
\begin{eqnarray} 
(\b{P}_{c,\l})_{ia,jb} = \sum_n \left( \b{X}_{n,\l}+\b{Y}_{n,\l} \right)_{ia} \left( \b{X}_{n,\l}+\b{Y}_{n,\l} \right)_{jb} -\delta_{ij}\delta_{ab},
\nonumber\\
\end{eqnarray}
or, in matrix form,
\begin{eqnarray} 
\b{P}_{c,\l} = \sum_n \left( \b{X}_{n,\l}+\b{Y}_{n,\l} \right) \left( \b{X}_{n,\l}+\b{Y}_{n,\l} \right)^\T -\b{1}.
\end{eqnarray}
Using the well-known fact that, if $\b{A}_\l+\b{B}_\l$ and $\b{A}_\l-\b{B}_\l$ are positive definite, the non-Hermitian eigenvalue equation~(\ref{eigeneq}) with real spin orbitals can be transformed into the following half-size symmetric eigenvalue equation
\begin{eqnarray} 
\b{M}_{\l} \b{Z}_{n,\l} = \omega_{n,\l}^2 \b{Z}_{n,\l},
\end{eqnarray}
where $\b{M}_{\l}=\left( \b{A}_\l-\b{B}_\l \right)^{1/2} \left( \b{A}_\l+\b{B}_\l \right) \left( \b{A}_\l-\b{B}_\l \right)^{1/2}$ and with eigenvectors $\b{Z}_{n,\l} = \sqrt{\omega_{n,\l}} \left( \b{A}_\l-\b{B}_\l \right)^{-1/2} \left( \b{X}_{n,\l}+\b{Y}_{n,\l} \right)$, and using the spectral decomposition $\b{M}_\l^{-1/2} = \sum_n \omega_{n,\l}^{-1} \b{Z}_{n,\l} \b{Z}_{n,\l}^\T$, the correlation two-particle density matrix $\b{P}_{c,\l}$ can be expressed as
\begin{eqnarray} 
\b{P}_{c,\l} = \left( \b{A}_\l-\b{B}_\l \right)^{1/2} \b{M}_\l^{-1/2} \left( \b{A}_\l-\b{B}_\l \right)^{1/2} -\b{1}.
\end{eqnarray}

\subsection{Expressions for spin-restricted closed-shell calculations}

For spin-restricted closed-shell calculations, the eigenvectors $(\b{X}_{n,\l}, \b{Y}_{n,\l})$ can be transformed into spin-singlet excitation/diexcitation vectors
\begin{subequations}
\begin{eqnarray} 
(^1\b{x}_{n,\l})_{ia}= \frac{1}{\sqrt{2}} \left[ \left( \b{X}_{n,\l} \right)_{i\uparrow a\uparrow} + \left( \b{X}_{n,\l} \right)_{i\downarrow a\downarrow} \right],
\end{eqnarray}
\begin{eqnarray} 
(^1\b{y}_{n,\l})_{ia}= \frac{1}{\sqrt{2}} \left[ \left( \b{Y}_{n,\l} \right)_{i\uparrow a\uparrow} + \left( \b{Y}_{n,\l} \right)_{i\downarrow a\downarrow} \right],
\end{eqnarray}
\end{subequations}
and spin-triplet excitation/diexcitation vectors
\begin{subequations}
\begin{eqnarray} 
(^{3,0}\b{x}_{n,\l})_{ia}= \frac{1}{\sqrt{2}} \left[ \left( \b{X}_{n,\l} \right)_{i\uparrow a\uparrow} - \left( \b{X}_{n,\l} \right)_{i\downarrow a\downarrow} \right],
\end{eqnarray}
\begin{eqnarray} 
(^{3,0}\b{y}_{n,\l})_{ia}= \frac{1}{\sqrt{2}} \left[ \left( \b{Y}_{n,\l} \right)_{i\uparrow a\uparrow} - \left( \b{Y}_{n,\l} \right)_{i\downarrow a\downarrow} \right],
\end{eqnarray}
\begin{eqnarray} 
(^{3,-1}\b{x}_{n,\l})_{ia}= \left( \b{X}_{n,\l} \right)_{i\uparrow a\downarrow},
\end{eqnarray}
\begin{eqnarray} 
(^{3,-1}\b{y}_{n,\l})_{ia}= \left( \b{Y}_{n,\l} \right)_{i\downarrow a\uparrow},
\end{eqnarray}
\begin{eqnarray} 
(^{3,1}\b{x}_{n,\l})_{ia}= \left( \b{X}_{n,\l} \right)_{i\downarrow a\uparrow},
\end{eqnarray}
\begin{eqnarray} 
(^{3,1}\b{y}_{n,\l})_{ia}= \left( \b{Y}_{n,\l} \right)_{i\uparrow a\downarrow},
\end{eqnarray}
\end{subequations}
the indices $i$, $a$, $j$, $b$ referring now to spatial orbitals.
With this transformation, the linear response eigenvalue equation~(\ref{eigeneq}) decouples into a singlet eigenvalue equation
\begin{eqnarray}
\left( \begin{array}{cc} ^1\b{A}_\l & ^1\b{B}_\l \\  ^1\b{B}_\l^* & ^1\b{A}_\l^* \\ \end{array} \right) \left( \begin{array}{c} ^1\b{x}_{n,\l} \\ ^1\b{y}_{n,\l} \\ \end{array} \right) =  {^1\omega_{n,\l}}  \left( \begin{array}{cc} \b{1} & \b{0}\\  \b{0} & -\b{1}\\ \end{array} \right) \left( \begin{array}{c} ^1\b{x}_{n,\l} \\ ^1\b{y}_{n,\l} \\ \end{array} \right),
\label{eigeneq1}
\end{eqnarray}
with the singlet orbital rotation Hessians
\begin{subequations}
\begin{eqnarray}
\left( ^1\b{A}_\l \right)_{ia,jb} &=&  (\epsilon_a-\epsilon_i) \delta_{ij} \delta_{ab} 
\nonumber\\
&&+ \l \left [ 2 \bra{i b} \hat{w}_{ee} \ket{a j} - \xi \bra{i b} \hat{w}_{ee} \ket{j a} \right],
\label{}
\end{eqnarray}
\begin{eqnarray}
\left( ^1\b{B}_\l \right)_{ia,jb} =  \l \left [ 2 \bra{a b} \hat{w}_{ee} \ket{i j} - \xi \bra{a b} \hat{w}_{ee} \ket{j i} \right],
\label{}
\end{eqnarray}
\end{subequations}
and three identical triplet eigenvalue equations
\begin{eqnarray}
\left( \begin{array}{cc} ^3\b{A}_\l & ^3\b{B}_\l \\  ^3\b{B}_\l^* & ^3\b{A}_\l^* \\ \end{array} \right) \left( \begin{array}{c} ^3\b{x}_{n,\l} \\ ^3\b{y}_{n,\l} \\ \end{array} \right) =  {^3\omega_{n,\l}}  \left( \begin{array}{cc} \b{1} & \b{0}\\  \b{0} & -\b{1}\\ \end{array} \right) \left( \begin{array}{c} ^3\b{x}_{n,\l} \\ ^3\b{y}_{n,\l} \\ \end{array} \right),
\label{eigeneq3}
\end{eqnarray}
with the triplet orbital rotation Hessians
\begin{subequations}
\begin{eqnarray}
\left( ^3\b{A}_\l \right)_{ia,jb} &=&  (\epsilon_a-\epsilon_i) \delta_{ij} \delta_{ab} - \l \xi \bra{i b} \hat{w}_{ee} \ket{j a},
\label{}
\end{eqnarray}
\begin{eqnarray}
\left( ^3\b{B}_\l \right)_{ia,jb} =  - \l \xi \bra{a b} \hat{w}_{ee} \ket{j i}.
\label{}
\end{eqnarray}
\end{subequations}
Performing the sums over spins in the correlation energy expression of Eq.~(\ref{Ecspinorbitals}), one gets, for real spatial orbitals,
\begin{eqnarray} 
E_c &=& \frac{1}{2} \int_{0}^{1} d\l \sum_{ia,jb} \bra{i b} \hat{w}_{ee} \ket{a j} (^1\b{P}_{c,\l})_{ia,jb},
\label{}
\end{eqnarray}
where remains only the contribution from the spin-singlet-adapted correlation two-particle density matrix $(^1\b{P}_{c,\l})_{ia,jb} = \sum_{\sigma_1=\uparrow,\downarrow} \sum_{\sigma_2=\uparrow,\downarrow} (\b{P}_{c,\l})_{i\sigma_1 a\sigma_1,j \sigma_2 b\sigma_2}$, which can be calculated similarly as before
\begin{eqnarray} 
^1\b{P}_{c,\l} &=& 2 \left[ \sum_n \left( {^1}\b{x}_{n,\l}+ {^1}\b{y}_{n,\l} \right) \left( {^1}\b{x}_{n,\l}+ {^1}\b{y}_{n,\l} \right)^\T -\b{1} \right]
\nonumber\\
               &=& 2 \left[ \left( {^1}\b{A}_\l-{^1}\b{B}_\l \right)^{1/2} {^1}\b{M}_\l^{-1/2} \left( {^1}\b{A}_\l-{^1}\b{B}_\l \right)^{1/2} -\b{1} \right],
\nonumber\\
\end{eqnarray}
where ${^1}\b{M}_{\l}=\left( {^1}\b{A}_\l-{^1}\b{B}_\l \right)^{1/2} \left( {^1}\b{A}_\l+{^1}\b{B}_\l \right) \left( {^1}\b{A}_\l-{^1}\b{B}_\l \right)^{1/2}$.

\bibliographystyle{apsrev}
\bibliography{biblio}

\begin{thebibliography}{79}
\bibitem[{Tou({\natexlab{a}})}]{TouZhuAngSav-JJJ-XX-note}
\bibinfo{note}{The short-range self-energy correction $\Delta \Sigma^\sr_\l$ is
  wrongly missing in Eq.~(11) of Ref.~\onlinecite{TouGerJanSavAng-PRL-09}.
  However, in practice, this term vanishes in the RPA or RPAx approximation so
  that the results of Ref.~\onlinecite{TouGerJanSavAng-PRL-09} are correct.}
\bibitem[{Tou({\natexlab{b}})}]{TouZhuAngSav-JJJ-XX-note2}
\bibinfo{note}{In the context of density-functional theory RPA is usually
  derived from the Kohn-Sham reference, while in the context of many-body
  perturbation theory (see appendices) RPA is usually derived from the HF
  reference. Therefore, both PBE+RPA and HF+RPA are theoretically justified.}
\bibitem[{Tou({\natexlab{c}})}]{TouZhuAngSav-JJJ-XX-note3}
\bibinfo{note}{The inverse of a 4-point function $\chi (1,2;1',2')$ is defined
  according to $\int d1'd2' \chi (1,2;1',2') \chi^{-1} (2',1';4,3) =
  \delta(1,3) \delta (2,4)$.}
\end{thebibliography}
\end{document}